\documentclass[10pt,a4paper]{article}
\usepackage{graphicx}
\usepackage{multicol}
\usepackage{tabto}
\usepackage{url}
\usepackage{enumitem}
\usepackage{textcomp}
\usepackage{eurosym}
\usepackage[labelfont=bf]{caption}
\usepackage[yyyymmdd]{datetime}

\usepackage{flowchart}
\usetikzlibrary{shapes,arrows.meta,chains,external}

\newcommand{\eat}[1]{}
\newcommand{\cz}[1]{\textit{\textbf{#1}}}

\newcommand{\ts}{
    \tikzset{>={Latex[width=2mm,length=2mm]}}
    \tikzstyle{line} = [draw, ->, >=latex, ultra thick]
    \tikzstyle{circ} = [
        circle,
        align=center,
        text width=2em,
        text centered,
        inner sep=1mm,
        outer sep=1mm,
        minimum width=0cm,
        minimum height=0cm
    ]
    \tikzstyle{tc} = [
      circle,
      draw,
      ultra thick,
      align=center,
      text width=2em,
      text centered,
      inner sep=1mm,
      outer sep=1mm,
      minimum width=0cm,
      minimum height=0cm
    ]
    \tikzstyle{csm} = [
        circle,
        align=center,
        text width=2em,
        text centered,
        inner sep=0mm,
        outer sep=0mm,
        minimum width=0cm,
        minimum height=0cm
    ]
    \tikzstyle{block} = [
        rectangle,
        draw,
        text centered,
        align=center,
        minimum width=10em,
        minimum height=10em
    ]
    \tikzstyle{noshape} = [text width=5em, text centered, minimum height=5em]
}

\begin{document}

\bibliographystyle{plain}
\thispagestyle{empty}

\NumTabs{6}

\begin{center}
\large{\bf A Decentralised Digital Identity Architecture}\\
\end{center}
\begin{center}
\begin{minipage}[t][][t]{0.44\linewidth}
\begin{center}
\large{\bf Geoff Goodell}\\
Centre for Blockchain Technologies\\
University College London\\
\texttt{g.goodell@ucl.ac.uk}
\end{center}
\end{minipage}
\begin{minipage}[t][][t]{0.44\linewidth}
\begin{center}
\large{\bf Tomaso Aste}\\
Centre for Blockchain Technologies\\
University College London
\texttt{t.aste@ucl.ac.uk}
\end{center}
\end{minipage}
\end{center}

\begin{abstract}

Current architectures to validate, certify, and manage identity are based on
centralised, top-down approaches that rely on trusted authorities and
third-party operators.  We approach the problem of digital identity starting
from a human rights perspective, with a primary focus on identity systems in
the developed world.  We assert that individual persons must be allowed to
manage their personal information in a multitude of different ways in different
contexts and that to do so, each individual must be able to create multiple
unrelated identities.  Therefore, we first define a set of fundamental
constraints that digital identity systems must satisfy to preserve and promote
privacy as required for individual autonomy.  With these constraints in mind,
we then propose a decentralised, standards-based approach, using a combination
of distributed ledger technology and thoughtful regulation, to facilitate
many-to-many relationships among providers of key services.  Our proposal for
digital identity differs from others in its approach to trust in that we do not
seek to bind credentials to each other or to a mutually trusted authority to
achieve strong non-transferability.  Because the system does not implicitly
encourage its users to maintain a single aggregated identity that can
potentially be constrained or reconstructed against their interests,
individuals and organisations are free to embrace the system and share in its
benefits.

\end{abstract}

\section{Introduction and Scope}
\label{s:i}

The past decade has seen a proliferation of new initiatives to create digital
identities for natural persons.  Some of these initiatives, such as the ID4D
project sponsored by the World Bank~\cite{id4d} and the Rohingya
Project~\cite{rohingya} involve a particular focus in the humanitarian context,
while others, such as Evernym~\cite{evernym} and ID2020~\cite{id2020} have a
more general scope that includes identity solutions for the developed world.
Some projects are specifically concerned with the rights of
children~\cite{5rights}.  Some projects use biometrics, which raise certain
ethical concerns~\cite{pandya2019}.  Some projects seek strong
non-transferability, either by linking all credentials related to a particular
natural person to a specific identifier, to biometric data, or to each other,
as is the case for the anonymous credentials proposed by Camenisch and
Lysyanskaya~\cite{camenisch2001}.  Some projects have design objectives that
include exceptional access (``backdoors'') for authorities, which are widely
considered to be problematic~\cite{abelson1997,abelson2015,benaloh2018}.

Although this article shall focus on challenges related to identity systems for
adult persons in the developed world, we argue that the considerations around
data protection and personal data that are applicable in the humanitarian
context, such as those elaborated by the International Committee of the Red
Cross~\cite{kuner2017} and Stevens~\cite{stevens2018}, also apply to the
general case.  We specifically consider the increasingly commonplace
application of identity systems ``to facilitate targeting, profiling and
surveillance'' by ``binding us to our recorded characteristics and
behaviours''~\cite{privacy-identity}.  Although we focus primarily upon the
application of systems for digital credentials to citizens of relatively
wealthy societies, we hope that our proposed architecture might contribute to
the identity zeitgeist in contexts such as humanitarian aid, disaster relief,
refugee migration, and the special interests of children as well.

We argue that while requiring strong non-transferability might be appropriate
for some applications, it is inappropriate and dangerous in others.
Specifically, we consider the threat posed by mass surveillance of ordinary
persons based on their habits, attributes, and transactions in the world.
Although the governments of Western democracies might be responsible for some
forms of mass surveillance, for example via the recommendations of the
Financial Action Task Force~\cite{fatf-recommendations} or various efforts to
monitor Internet activity~\cite{snoopers,australia2018}, the siren song of
surveillance capitalism~\cite{zuboff2015}, including the practice of ``entity
resolution'' through aggregation and data analysis~\cite{waldman}, presents a
particular risk to human autonomy.

We suggest that many ``everyday'' activities such as the use of library
resources, public transportation services, and mobile data services are
included in a category of activities for which strong non-transferability is
not necessary and for which there is a genuine need for technology that
explicitly protects the legitimate privacy interests of individual persons.  We
argue that systems that encourage individual persons to establish a single,
\textit{unitary}\footnote{In the context of personal identity, use the term
\textit{unitary} to refer to attributes, transactions, or identifiers for which
an individual can have at most one and that are, for practical purposes,
inseparably bound to their subject.} avatar (or ``master key'') for use in many
contexts can ultimately influence and constrain how such persons behave, and we
suggest that if a link between two attributes or transactions can be proven,
then it can be forcibly discovered.  We therefore argue that support for
multiple, unlinkable identities is an essential right and a necessity for the
development of a future digital society for humans.

This rest of this article is organised as follows.  In the next section
section, we offer some background on identity systems; we frame the problem
space and provide examples of existing solutions.  In Section 3, we introduce a
set of constraints that serve as properties that a digital identity
infrastructure must have to support human rights.  In Section 4, we describe
how a digital identity system with a fixed set of actors might operate and how
it might be improved.  In Section 5, we introduce distributed ledger technology
to promote a competitive marketplace for issuers and verifiers of credentials
and to constrain the interaction between participants in a way that protects
the privacy of individual users.  In Section 6, we consider how the system
should be operated and maintained if it is to satisfy the human rights
requirements.  In Section 7, we suggest some potential use cases, and in
Section 8 we conclude.

\section{Background}
\label{s:m}

Establishing meaningful credentials for individuals and organisations in an
environment in which the various authorities are not uniformly trustworthy
presents a problem for currently deployed services, which are often based on
hierarchical trust networks, all-purpose identity cards, and other artifacts of
the surveillance economy.  In the context of interactions between natural
persons, identities are neither universal nor hierarchical, and a top-down
approach to identity generally assumes that it is possible to impose a
universal hierarchy.  Consider ``Zooko's triangle,''  which states that names
can be distributed, secure, or human-readable, but not all three~\cite{zooko}.
The stage names of artists may be distributed and human-readable but are not
really secure since they rely upon trusted authorities to resolve conflicts.
The names that an individual assigns to friends or that a small community
assigns to its members (``petnames''~\cite{petnames}) are secure and
human-readable but not distributed.  We extend the reasoning behind the paradox
to the problem of identity itself and assert that the search for unitary
identities for individual persons is problematic.  It is technically
problematic because there is no endogenous way to ensure that an individual has
only one self-certifying name~\cite{douceur}, there is no way to be sure about
the trustworthiness or universality of an assigned name, and there is no way to
ensure that an individual exists only within one specific community.  More
importantly, we assert that the ability to manage one's identities in a
multitude of different contexts, including the creation of multiple unrelated
identities, is an essential human right.

\subsection{Manufacturing Trust}

The current state-of-the-art identity systems, from technology platforms to
bank cards, impose asymmetric trust relationships and contracts of adhesion on
their users, including both the ultimate users as well as local authorities,
businesses, cooperatives, and community groups.  Such trust relationships,
often take the form of a hierarchical trust infrastructure, requiring that
users accept either a particular set of trusted certification authorities
(``trust anchors'') or identity cards with private keys generated by a trusted
third party.  In such cases, the systems are susceptible to socially
destructive business practices, corrupt or unscrupulous operators, poor
security practices, or control points that risk coercion by politically or
economically powerful actors.  Ultimately, the problem lies in the dubious
assumption that some particular party or set of parties are universally
considered trustworthy.

Often, asymmetric trust relationships set the stage for security breaches.
Rogue certification authorities constitute a well-known risk, even to
sophisticated government actors~\cite{vanderburg,charette}, and forged
signatures have been responsible for a range of cyber-attacks including the
Stuxnet worm, an alleged cyber-weapon believed to have caused damage to Iran's
nuclear programme~\cite{stuxnet}, as well as a potential response to Stuxnet by
the government of Iran~\cite{eckersley}.  Corporations that operate the largest
trust anchors have proven to be vulnerable.  Forged credentials were
responsible for the Symantec data breach~\cite{arstechnica3}, and other popular
trust anchors such as Equifax are not immune to security
breaches~\cite{equifax}.  Google has published a list of certification
authorities that it thinks are untrustworthy~\cite{register}, and IT
administrators have at times undermined the trust model that relies upon root
certification authorities~\cite{slashdot}.  Finally, even if their systems are
secure and their operators are upstanding, trust anchors are only as secure as
their ability to resist coercion, and they are sometimes misappropriated by
governments~\cite{bright}.

Such problems are global, affecting the developed world and emerging economies
alike.  Identity systems that rely upon a single technology, a single
implementation, or a single set of operators have proven
unreliable~\cite{arstechnica1,arstechnica2,engadget1}.  Widely-acclaimed
national identity systems, including but not limited to the Estonian identity
card system based on X-Road~\cite{estonia} and Aadhaar in India~\cite{aadhaar},
are characterised by centralised control points, security risks, and
surveillance.


Recent trends in technology and consumer services suggest that concerns about
mobility and scalability will lead to the deployment of systems for identity
management that identify consumers across a variety of different services, with
a new marketplace for providers of identification services~\cite{wagner2014}.
In general, the reuse of credentials has important privacy implications as a
consumer's activities may be tracked across multiple services or multiple uses
of the same service.   For this reason, the potential for a system to collect
and aggregate transaction data must be evaluated whilst evaluating its impact
on the privacy of its users.

While data analytics are becoming increasingly effective in identifying and
linking the digital trails of individual persons, it has become correspondingly
necessary to defend the privacy of individual users and implement instruments
that allow and facilitate anonymous access to services.  This reality was
recognised by the government of the United Kingdom in the design of its GOV.UK
Verify programme~\cite{govukverify}, a federated network of identity providers
and services.  However, the system as deployed has significant technical
shortcomings with the potential to jeopardise the privacy of its
users~\cite{brandao2015,whitley2018}, including a central hub and
vulnerabilities that can be exploited to link individuals with the services
they use~\cite{ohara2011}.

Unfortunately, not only do many of the recently-designed systems furnish or
reveal data about their users against their interests, but they have been
explicitly designed to do so.  For example, consider digital rights management
systems that force users to identify themselves \textit{ex ante} and then use
digital watermarks to reveal their identities~\cite{thomas2009}.  In some
cases, demonstrable privacy has been considered an undesirable feature and
designs that protect the user's identity intrinsically are explicitly excluded,
for example in the case of vehicular ad-hoc networks~\cite{shuhaimi2012}, with
the implication that systems without exceptional access features are dangerous.
Finally, of particular concern are systems that rely upon biometrics for
identification.  By binding identification to a characteristic that users (and
in most cases even governments) cannot change, biometrics implicitly prevent a
user from transacting within a system without connecting each transaction to
each other and potentially to a permanent record.  In recent years, a variety
of US patents have been filed and granted for general-purpose identity systems
that rely upon biometric data to create a ``root'' identity linking all
transactions in this manner~\cite{liu2008,thackston2018}.

\subsection{Approaches Using Distributed Ledgers}
\label{ss:da}

\begin{figure}
\begin{center}
\scalebox{0.9}{\begin{tikzpicture}[>=latex, node distance=3cm, font={\sf \small}, auto]\ts
\node (r1) at (-1.1,2) [tc, outer sep=0mm, text width=1.2em] {};
\node (r2) at (1.1,2) [tc, outer sep=0mm, text width=1.2em] {};
\node (r3) at (-2.2,0) [tc,  outer sep=0mm,text width=1.2em] {};
\node (r7) at (0,0) [tc, outer sep=0mm, text width=5.6em, fill=gray!20] {};
\node (t7) at (0,0) [circ, text width=4.8em, fill=gray!20] {Platform\\Operator};
\node (r4) at (2.2,0) [tc, outer sep=0mm,text width=1.2em] {};
\node (r5) at (-1.1,-2) [tc, outer sep=0mm, text width=1.2em] {};
\node (r6) at (1.1,-2) [tc, outer sep=0mm, text width=1.2em] {};
\tikzset{>={Latex[width=3mm,length=3mm]}}
\draw[<->, line width=1.0mm] (r1) -- (r7);
\draw[<->, line width=1.0mm] (r2) -- (r7);
\draw[<->, line width=1.0mm] (r3) -- (r7);
\draw[<->, line width=1.0mm] (r4) -- (r7);
\draw[<->, line width=1.0mm] (r5) -- (r7);
\draw[<->, line width=1.0mm] (r6) -- (r7);
\end{tikzpicture}}
\scalebox{0.9}{\begin{tikzpicture}[>=latex, node distance=3cm, font={\sf \small}, auto]\ts
\node (r1) at (-1.1,2) [tc, outer sep=0mm, text width=1.2em] {};
\node (r2) at (1.1,2) [tc, outer sep=0mm, text width=1.2em] {};
\coordinate (c1) at (-5,0);
\coordinate (c2) at (-2.75,0);
\node (r3) at (-2.2,0) [tc, outer sep=0mm, text width=1.2em] {};
\node (r7) at (0,0) [tc, outer sep=0mm, text width=8.4em, dashed] {Distributed\\Ledger};
\node (r4) at (2.2,0) [tc, outer sep=0mm, text width=1.2em] {};
\node (r5) at (-1.1,-2) [tc, outer sep=0mm, text width=1.2em] {};
\node (r6) at (1.1,-2) [tc, outer sep=0mm, text width=1.2em] {};
\tikzset{>={Latex[width=6mm,length=6mm]}}
\draw[->, line width=2.0mm] (c1) -- (c2);
\tikzset{>={Latex[width=2mm,length=2mm]}}
\draw[-, line width=1.0mm] (r1) -- (r7);
\draw[-, line width=1.0mm] (r2) -- (r7);
\draw[-, line width=1.0mm] (r3) -- (r7);
\draw[-, line width=1.0mm] (r4) -- (r7);
\draw[-, line width=1.0mm] (r5) -- (r7);
\draw[-, line width=1.0mm] (r6) -- (r7);
\end{tikzpicture}}

\caption{\textit{Many network services are centralised in the sense that
participants rely upon a specific platform operator to make use of the service
(left), whereas distributed ledgers rely upon network consensus among
participants instead of platform operators (right).}}

\label{f:platform}
\end{center}
\end{figure}
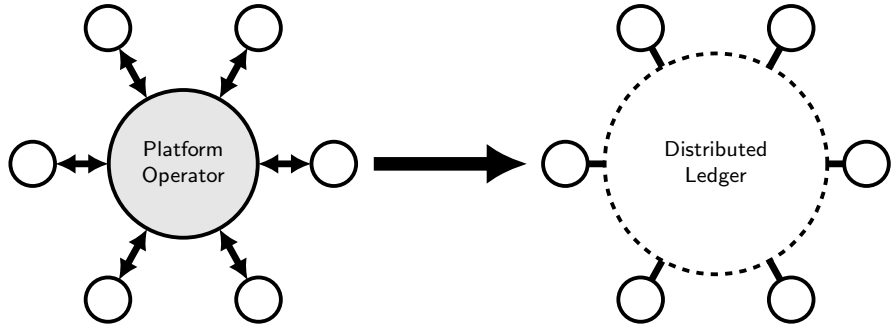

The prevailing identity systems commonly require users to accept third parties
as trustworthy.  The alternative to imposing new trust relationships is to work
with existing trust relationships by allowing users, businesses, and
communities to deploy technology on their own terms, independently of external
service providers.  In this section we identify various groups that have
adopted a system-level approach to allow existing institutions and service
providers to retain their relative authority and decision-making power without
forcibly requiring them to cooperate with central authorities (such as
governments and institutions), service providers (such as system operators), or
the implementors of core technology.  We suggest that ideally, a solution would
not require existing institutions and service providers to operate their own
infrastructure without relying upon a platform operator, while concordantly
allowing groups such as governments and consultants to act as advisors,
regulators, and auditors, not operators.  A distributed ledger can serve this
purpose by acting as a neutral conduit among its participants, subject to
governance limitations to ensuring neutrality and design limitations around
services beyond the operation of the ledger that are required by participants.
Figure~\ref{f:platform} offers an illustration.

Modern identity systems are used to coordinate three activities:
identification, authentication, and authorisation.  The central problem to
address is how to manage those functions in a decentralised context with no
universally trusted authorities.  Rather than trying to force all participants
to use a specific new technology or platform, we suggest using a
multi-stakeholder process to develop common standards that define a set of
rules for interaction.  Any organisation would be able to develop and use their
own systems that would interoperate with those developed by any other
organisation without seeking permission from any particular authority or
agreeing to deploy any particular technology.


A variety of practitioners have recently proposed using a distributed ledger to
decentralise the administration of an identity system~\cite{dunphy2018}, and we
agree that the properties of distributed ledger technologies are appropriate
for the task.  In particular, distributed ledgers allow their participants to
share control of the system.  They also provide a common view of transactions
ensuring that everyone sees the same transaction history.

Various groups have argued that distributed ledgers might be used to mitigate
the risk that one powerful, central actor might seize control under the mantle
of operational efficiency.  However, it is less clear that this lofty goal is
achieved in practice.  Existing examples of DLT-enabled identity management
systems backed by organisations include the following, among others:

\begin{itemize}

\item\textit{ShoCard}~\cite{shocard} is operated by a commercial entity that serves
as a trusted intermediary~\cite{dunphy2018}.

\item\textit{Everest}~\cite{everest} is designed as a payment solution backed
by biometric identity for its users.  The firm behind Everest manages the
biometric data and implicitly requires natural persons to have at most one
identity within the system~\cite{graglia2018}.

\item\textit{Evernym}~\cite{evernym} relies on a foundation
(Sovrin~\cite{sovrin}) to manage the set of approved certification
authorities~\cite{aitken2018}, and whether the foundation could manage the
authorities with equanimity remains to be tested.

\item\textit{ID2020}~\cite{id2020} offers portable identity using biometrics to
achieve strong non-transferability and persistence~\cite{prnewswire2019}.

\item\textit{uPort}~\cite{uport} does not rely upon a central authority,
instead allowing for mechanisms such as social recovery.  However, its design
features an optional central registry that might introduce a means of linking
together transactions that users would prefer to keep
separate~\cite{dunphy2018}.  The uPort architecture is linked to phone numbers
and implicitly discourages individuals from having multiple identities within
the system~\cite{graglia2018}.

\end{itemize}

\begin{figure}
\begin{center}
\scalebox{0.9}{\begin{tikzpicture}[>=latex, node distance=3cm, font={\sf \small}, auto]\ts
\node (r1) at (-1.1,2) [tc, text width=5.2em, fill=cyan!20] {};
\node (t1) at (-1.1,2) [circ, text width=5.2em] {Passport\\Office};
\node (r2) at (1.1,2) [tc, text width=5.2em, fill=blue!20] {};
\node (t2) at (1.1,2) [circ, text width=5.2em] {Vehicle\\Registry};
\node (r3) at (-2.2,0) [tc, text width=5.2em, fill=green!20] {};
\node (t3) at (-2.2,0) [circ, text width=5.2em] {Bank};
\node (r4) at (2.2,0) [tc, text width=5.2em, fill=magenta!20] {};
\node (t4) at (2.2,0) [circ, text width=5.2em] {Social\\Media};
\node (r5) at (-1.1,-2) [tc, text width=5.2em, fill=yellow!20] {};
\node (t5) at (-1.1,-2) [circ, text width=5.2em] {Email\\Service};
\node (r6) at (1.1,-2) [tc, text width=5.2em, fill=red!20] {};
\node (t6) at (1.1,-2) [circ, text width=5.2em] {Mobile\\Carrier};
\end{tikzpicture}}
\scalebox{0.9}{\begin{tikzpicture}[>=latex, node distance=3cm, font={\sf \small}, auto]\ts
\node (r1) at (-1.1,2) [tc, outer sep=0mm, text width=5.2em, fill=cyan!20] {};
\node (t1) at (-1.1,2) [circ, text width=5.2em] {Passport\\Office};
\node (r2) at (1.1,2) [tc, outer sep=0mm, text width=5.2em, fill=blue!20] {};
\node (t2) at (1.1,2) [circ, text width=5.2em] {Vehicle\\Registry};
\coordinate (c) at (-5,0);
\node (r3) at (-2.2,0) [tc, outer sep=0mm, text width=5.2em, fill=green!20] {};
\node (t3) at (-2.2,0) [circ, text width=5.2em] {Bank};
\node (r7) at (0,0) [tc, outer sep=0mm, text width=5.2em, fill=gray!20] {};
\node (t7) at (0,0) [circ, text width=5.2em] {Unitary Avatar};
\node (r4) at (2.2,0) [tc, outer sep=0mm, text width=5.2em, fill=magenta!20] {};
\node (t4) at (2.2,0) [circ, text width=5.2em] {Social\\Media};
\node (r5) at (-1.1,-2) [tc, outer sep=0mm, text width=5.2em, fill=yellow!20] {};
\node (t5) at (-1.1,-2) [circ, text width=5.2em] {Email\\Service};
\node (r6) at (1.1,-2) [tc, outer sep=0mm, text width=5.2em, fill=red!20] {};
\node (t6) at (1.1,-2) [circ, text width=5.2em] {Mobile\\Carrier};
\tikzset{>={Latex[width=6mm,length=6mm]}}
\draw[->, line width=2.0mm] (c) -- (t3);
\tikzset{>={Latex[width=2mm,length=2mm]}}
\draw[-, line width=1.0mm] (r1) -- (r7);
\draw[-, line width=1.0mm] (r2) -- (r7);
\draw[-, line width=1.0mm] (r3) -- (r7);
\draw[-, line width=1.0mm] (r4) -- (r7);
\draw[-, line width=1.0mm] (r5) -- (r7);
\draw[-, line width=1.0mm] (r6) -- (r7);
\end{tikzpicture}}

\caption{\textit{Consider that individual persons possess credentials
representing a variety of attributes (left), and schemes that attempt to achieve
strong non-transferability seek to bind these attributes together into a
single, unitary ``avatar'' or ``master'' identity (right).}}

\label{f:unitary}
\end{center}
\end{figure}
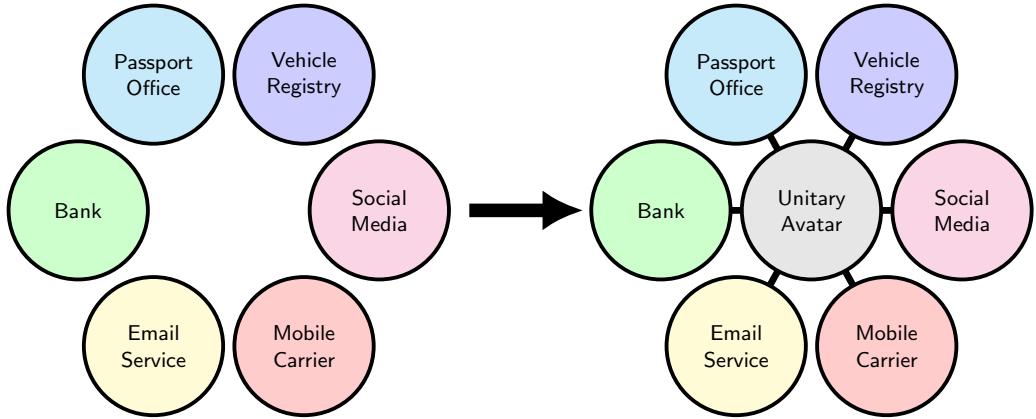

Researchers have proposed alternative designs to address some of the concerns.
A design suggested by Kaaniche and Laurent does not require a central authority
for its blockchain infrastructure but does require a trusted central entity for
its key infrastructure~\cite{kaaniche2017}.  Coconut, the selective disclosure
credential scheme used by Chainspace~\cite{chainspace}, is designed to be
robust against malicious authorities and may be deployed in a way that
addresses such concerns~\cite{sonnino2018}.\footnote{Chainspace was acquired by
Facebook in early 2019, and its core technology subsequently became central to
the Libra platform~\cite{heath2019,field2019}.}  We find that many systems such
as these require users to bind together their credentials \textit{ex
ante}\footnote{We use the term \textit{early-binding} to refer to systems that
establish provable relationships between transactions, attributes, identifiers,
or credentials before they are used.  We use the term \textit{late-binding} to
refer to systems that allow their users to establish such relationships at the
time of use.} to achieve non-transferability, essentially following a design
proposed by Camenisch and Lysyanskaya~\cite{camenisch2001} that establishes a
single ``master key'' that allows each user to prove that all of her
credentials are related to each other.  Figure~\ref{f:unitary} offers an
illustration.  Even if users were to have the option to establish multiple
independent master keys, service providers or others could undermine that
option by requiring proof of the links among their credentials.

The concept of an individual having ``multiple identities'' is potentially
confusing, so let us be clear.  In the context of physical documents in the
developed world, natural persons generally possess multiple identity documents
already, including but not limited to passports, driving licenses, birth
certificates, bank cards, insurance cards, and so on.  Although individuals
might not think of these documents and the attributes they represent as
constituting multiple identities, the identity documents generally stand alone
for their individual, limited purposes and need not be presented as part of a
bundled set with explicit links between the attributes.  Service providers
might legitimately consider two different identity documents as pertaining to
two different individuals, even whilst they might have been issued to the same
person.  A system that links together multiple attributes via early-binding
eliminates this possibility.  When we refer to ``multiple identities'' we refer
to records of attributes or transactions that are not linked to each other.
Users of identity documents might be willing to sacrifice this aspect of
control in favour of convenience, but the potential for blacklisting and
surveillance that early-binding introduces is significant.  It is for this
reason that we take issue with the requirement, advised by various groups
including the International Telecommunications Union~\cite{itu2018}, that
individuals must not possess more than one identity.  Such a requirement is
neither innocuous nor neutral.

\begin{table}
\begin{center}\sf
\def\arraystretch{1.3}
\begin{tabular}{|p{2.9cm}p{4cm}p{7cm}|}\hline
Name & Objectives & Concerns \\ \hline
\multicolumn{3}{|l|}{\textbf{Government-operated solutions}} \\
Estonian ID-Card
    & interoperability, assurance
    & centralised governance, surveillance \\
Aadhaar (India)
    & interoperability, assurance
    & centralised governance, surveillance \\
GOV.UK Verify
    & federated management
    & central hub, surveillance \\
\hline\multicolumn{3}{|l|}{\textbf{Privately-operated solutions}} \\
ShoCard
    & \raggedright strong non-transferability, auditability
    & \raggedright commercial entity is a trusted intermediary, commercial entity stores biometric data \tabularnewline
Everest
    & \raggedright strong non-transferability, digital payments
    & commercial entity stores biometric data \\
ID2020
    & \raggedright portability, persistence, strong non-transferability
    & identities are unitary through use of biometrics \\
Evernym
    & federated management
    & private foundation has an operational role \\
Kaaniche/Laurent
    & hierarchical management
    & \raggedright requires an agreed-upon hierarchy with trusted authority \tabularnewline
\hline\multicolumn{3}{|l|}{\textbf{Decentralised architectures}} \\
uPort
    & \raggedright federated governance and management
    & \raggedright identities become unitary through early-binding or similar mechanisms \tabularnewline
Chainspace
    & \raggedright federated governance and management
    & \raggedright identities become unitary through early-binding or similar mechanisms \tabularnewline
\hline\end{tabular}
\caption{\textit{A characterisation of the landscape of digital identity solutions, with examples.}}
\label{t:landscape}
\rm\end{center}
\end{table}

Table~\ref{t:landscape} summarises the landscape of prevailing digital
identity solutions.  We imagine that the core technology underpinning these and
similar approaches might be adapted to implement a protocol that is broadly
compatible with what we describe in this article.  However, we suspect that in
practice they would need to be modified to encourage users to establish
multiple, completely independent identities.  In particular, service providers
would not be able to assume that users have bound their credentials to each
other \textit{ex ante}, and if non-transferability is required, then the system
would need to achieve it in a different way.

\subsection{Participants in an Identity System}

We shall use the following notation to represent the various parties that
interact with a typical identity system:

\begin{itemize}

\item (1) A \textit{``certification provider''} (\textbf{CP}).  This would be
an entity or organisation responsible for establishing a credential based upon
foundational data.  The credential can be used as a form of identity and
generally represents that the organisation has checked the personal identity of
the user in some way.  In the context of digital payments, this might be a
bank.

\item (2) An \textit{``authentication provider''} (\textbf{AP}).  This would be
any entity or organisation that might be trusted to verify that a credential is
valid and has not been revoked.  In current systems, this function is typically
performed by a platform or network, for example a payment network such as those
associated with credit cards.

\item (3) An \textit{``end-user service provider''} (\textbf{Service}).  This
would be a service that requires a user to provide credentials.  It might be a
merchant selling a product, a government service, or some other kind of
gatekeeper, for example a club or online forum.

\item (4) A \textit{user} (\textbf{user}).  This would be a human operator, in
most cases aided by a device or a machine, whether acting independently or
representing an organisation or business.

\end{itemize}

As an example of how this might work, suppose that a user wants to make an
appointment with a local consular office.  The consular office wants to know
that a user is domiciled in a particular region.  The user has a bank account
with a bank that is willing to certify that the user is domiciled in that
region.  In addition, a well-known authentication provider is willing to accept
certifications from the bank, and the consular office accepts signed statements
from that authentication provider.  Thus, the user can first ask the bank to
sign a statement certifying that he is domiciled in the region in question.
When the consular office asks for proof of domicile, the user can present the
signed statement from the bank to the authentication provider and ask the
authentication provider to sign a new statement vouching for the user's region
of domicile, using information from the bank as a basis for the statement,
without providing any information related to the bank to the consular office.

\section{Design Constraints for Privacy as a Human Right}
\label{s:constraints}

Reflecting on the various identity systems used today, including but not
limited to residence permits, bank accounts, payment cards, transit passes, and
online platform logins, we observed a plethora of features with weaknesses and
vulnerabilities concerning privacy (and in some cases security) that could
potentially infringe upon human rights.  Although the 1948 Universal
Declaration on Human Rights explicitly recognises privacy as a human
right~\cite{un1948}, the declaration was drafted well before the advent of a
broad recognition of the specific dangers posed by the widespread use of
computers for data aggregation and analysis~\cite{armer1975}, to say nothing of
surveillance capitalism~\cite{zuboff2015}.  Our argument that privacy in the
context of digital identity is a human right, therefore, rests upon a more
recent consideration of the human rights impact of the abuse of economic
information~\cite{stoa1999}.  With this in mind, we identified the following
eight fundamental constraints to frame our design requirements for technology
infrastructure~\cite{goodell18}:

\noindent{\bf \textit{Structural requirements:}}

\vspace{-10pt}
\begin{enumerate}
\setlength\itemsep{0.5em}

\item \textit{Minimise \cz{control points} that can be used to co-opt the
system.}  A single point of trust is a single point of failure, and both state
actors and technology firms have historically been proven to abuse such trust.

\item \textit{Resist establishing potentially abusive \cz{processes and
practices}, including legal processes, that rely upon control points.}
Infrastructure that can be used to abuse and control individual persons is
problematic even if those who oversee its establishment are genuinely benign.
Once infrastructure is created, it may in the future be used for other purposes
that benefit its operators.

\end{enumerate}

\vspace{-10pt}\noindent{\bf \textit{Human requirements:}}

\vspace{-10pt}
\begin{enumerate}[resume]
\setlength\itemsep{0.5em}

\item \textit{Mitigate architectural characteristics that lead to \cz{mass
surveillance} of individual persons.}  Mass surveillance is about control as
much as it is about discovery: people behave differently when they believe that
their activities are being monitored or evaluated~\cite{mayo45}.  Powerful
actors sometimes employ monitoring to create incentives for individual persons,
for example to conduct marketing promotions or credit scoring operations.  Such
incentives may prevent individuals from acting autonomously, and the chance to
discover irregularities, patterns, or even misbehaviour often does not justify
such mechanisms of control.

\item \textit{Do not impose \cz{non-consensual} trust relationships upon
beneficiaries.}  It is an act of coercion for a service provider to require a
client to maintain a direct trust relationship with a specific third-party
platform provider or certification authority.  Infrastructure providers must
not explicitly or implicitly engage in such coercion, which should be
recognised for what it is and not tolerated in the name of convenience.

\item \textit{Empower individual users to manage the \cz{linkages} among their
activities.}  To be truly free and autonomous, individuals must be able to
manage the cross sections of their activities, attributes, and transactions
that are seen or might be discovered by various institutions, businesses, and
state actors.

\end{enumerate}

\vspace{-10pt}\noindent{\bf \textit{Economic requirements:}}

\vspace{-10pt}
\begin{enumerate}[resume]
\setlength\itemsep{0.5em}

\item \textit{Prevent solution providers from establishing a \cz{monopoly
position}.}  Some business models are justified by the opportunity to achieve
status as monopoly infrastructure.  Monopoly infrastructure is problematic not
only because it deprives its users of consumer surplus but also because it
empowers the operator to dictate the terms by which the infrastructure can be
used.

\item \textit{Empower local businesses and cooperatives to establish their own
\cz{trust relationships}.}  The opportunity to establish trust relationships on
their own terms is important both for businesses to compete in a free
marketplace and for businesses to act in a manner that reflects the interests
of their communities.

\item \textit{Empower service providers to establish their own \cz{business
practices and methods}.}  Providers of key services must adopt practices that
work within the values and context of their communities.

\end{enumerate}

These constraints constitute a set of \textit{system-level} requirements,
involving human actors, technology, and their interaction, not to be confused
with the \textit{technical} requirements that have been characterised as
essential to self-sovereign identity (SSI)~\cite{stevens2018}.  Although our
design objectives may overlap with the design objectives for SSI systems, we
seek to focus on system-level outcomes.  While policy changes at the government
level might be needed to fully achieve the vision suggested by some of the
requirements, we would hope that a digital identity system would not contain
features that intrinsically facilitate their violation.

Experience shows that control points will eventually be co-opted by powerful
parties, irrespective of the intentions of those who build, own, or operate the
control points.  Consider, for example, how Cambridge Analytica allegedly
abused the data assets of Facebook Inc to manipulate voters in Britain and the
US~\cite{koslowska} and how the Russian government asserted its influence on
global businesses that engaged in domain-fronting~\cite{lunden,savov}.  The
inherent risk that centrally aggregated datasets may be abused, not only by the
parties doing the aggregating but also by third parties, implies value in
system design that avoids control points and trusted infrastructure operators,
particularly when personal data and livelihoods are involved.

\section{A Digital Identity System}

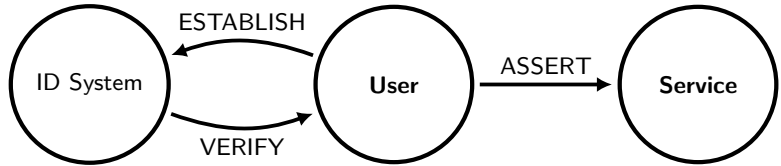
\begin{figure}
\begin{center}
\begin{tikzpicture}[>=latex, node distance=3cm, font={\sf \small}, auto]\ts
\node (r1) at (0,0) [circ, draw, ultra thick, text width=5.2em] {ID System};
\node (r2) at (4,0) [circ, draw, ultra thick, text width=5.2em] {\textbf{User}};
\node (r3) at (8,0) [circ, draw, ultra thick, text width=5.2em] {\textbf{Service}};

\draw[->, line width=0.5mm] (r2) edge[bend right=20] node[sloped, above] {ESTABLISH} (r1);
\draw[->, line width=0.5mm] (r1) edge[bend right=20] node[sloped, below] {VERIFY} (r2);
\draw[->, line width=0.5mm] (r2) -- node[sloped, above] {ASSERT} (r3);

\end{tikzpicture}

\caption{\textit{A schematic representation of a generalised identity system.
Users first establish a credential with the system, then use the system to
verify the credential, and then use the verified identity to assert that they
are authorised to receive a service.}}

\label{f:overview}
\end{center}
\end{figure}

Various digital identity architectures and deployments exist today to perform
the three distinct functions we mentioned earlier: identification,
authentication, and authorisation~\cite{riley}.  We introduce a fourth
function, \textit{auditing}, by which the basis for judgements made by the
system can be explained and evaluated.  We characterise the four functions as
follows:

\begin{itemize}

\item \textsc{Identification.} A user first \textit{establishes} some kind of
credential or identifier.  The credential might be a simple registration, for
example with an authority, institution, or other organisation.  In other cases,
it might imply a particular attribute.  The implication might be implicit, as a
passport might imply citizenship of a particular country or a credential issued
by a bank might imply a banking relationship, or it might be explicit, as in
the style of attribute-backed
credentials~\cite{ibm1,camenisch2003}.\footnote{We do not describe how to use
attribute-backed credentials here.}

\item \textsc{Authentication.} Next, when the provider of a service seeks to
authenticate a user, the user must be able to \textit{verify} that a
credential in question is valid.

\item \textsc{Authorisation.} Finally, the user can use the authenticated
credential to \textit{assert} to the service provider that she is entitled to a
particular service.

\item \textsc{Auditing.} The identity system would maintain record of the
establishment, expiration, and revocation of credentials such that the success
or failure of any given authentication request can be explained.

\end{itemize}

Ultimately, it is the governance of a digital identity system, including its
intrinsic policies and mechanisms as well as the accountability of the
individuals and groups who control its operation, that determines whether it
empowers or enslaves its users.  We suggest that proper governance,
specifically including a unified approach to the technologies and policies that
the system comprises, is essential to avoiding unintended consequences to its
implementation.  We address some of these issues further in
Section~\ref{s:governance}.

\begin{table}
\begin{center}
\begin{tabular}{|ll|}\hline
{\sf request} & a request for a credential.\\
{\sf request $x$} & a request for a credential with a parameter, $x$.\\
{\sf A(m)} & A message $m$ signed by party {\sf A}.\\
{\sf identify} & the foundational identifying elements that a user presents to a certification\\
& provider, encrypted so that other parties (including AP) cannot read them.\\
{\sf revoke-one} & a message invalidating an earlier signature on a specific user credential.\\
{\sf revoke-all} & a message invalidating all signatures by a certain key.\\
{\sf [m]} & blinded version of message $m$.\\
{\sf A([m])} & blind signature of $m$ by {\sf A}.\\
{\sf prove-owner $x^*$} & proof of ownership of some private key $x^*$, for example via a challenge-response\\
& (which would imply two extra messages not shown) or by using the key to sign a\\
& pre-existing secret created by the recipient and shared with the sender.\\
{\sf request-certs A} & a request for all of the certificates on the ledger signed by {\sf A},\\
& followed by a response containing all of the matching certificates.\\
{\sf receipt} & response from the distributed ledger system indicating that a transaction\\
& completed successfully.\\
{\sf object} & physical, tamper-resistant object containing a unique receipt for a transaction.\\
\hline\end{tabular}
\caption{\textit{Notation used in the subsequent figures.}}
\label{t:lexicon}
\end{center}
\end{table}

Figure~\ref{f:overview} gives a pictorial representation of the functions.
Table~\ref{t:lexicon} defines the notation that we shall use in our figures.
With the constraints enumerated in Section~\ref{s:constraints} taken as design
requirements, we propose a generalised architecture that achieves our
objectives for an identity system.  The candidate systems identified in
Section~\ref{s:m} can be evaluated by comparing their features to our
architecture.  Since we intend to argue for a practical solution, we start with
a system currently enjoying widespread deployment.

\subsection{SecureKey Concierge}
\label{ss:skc}

As a baseline example of an identity framework, we consider a system that uses
banks as certification providers whilst circumventing the global payment
networks.  \textit{SecureKey Concierge} (SKC)~\cite{skc} is a solution used by
the government of Canada to provide users with access to its various systems in
a standard way.  The SKC architecture seeks the following benefits:

\begin{enumerate}

\item Leverage existing ``certification providers'' such as banks and other
financial institutions with well-established, institutional procedures for
ascertaining the identities of their customers.  Often such procedures are
buttressed by legal frameworks such as Anti-Money Laundering (AML) and ``Know
Your Customer'' (KYC) regulations that broadly deputise banks and substantially
all financial institutions~\cite{govuk2014} to collect identifying information
on the various parties that make use of their services, establish the expected
pattern for the transactions that will take place over time, and monitor the
transactions for anomalous activity inconsistent with the
expectations~\cite{amlkyc}.

\item Isolate service providers from personally identifying bank details and
eliminate the need to share specific service-related details with the
certification provider, whilst avoiding traditional authentication service
providers such as payment networks.

\end{enumerate}

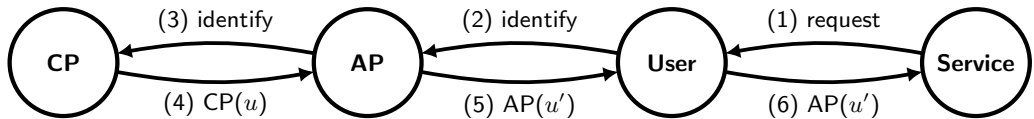
\begin{figure}
\begin{center}
\begin{tikzpicture}[>=latex, node distance=3cm, font={\sf \small}, auto]\ts

\node (ip) at (-4,0) [csm, draw, ultra thick, text width=4em] {\textbf{CP}};
\node (ap) at (0,0) [csm, draw, ultra thick, text width=4em] {\textbf{AP}};
\node (user) at (4,0) [csm, draw, ultra thick, text width=4em] {\textbf{User}};
\node (sp) at (8,0) [csm, draw, ultra thick, text width=4em] {\textbf{Service}};

\draw[->, line width=0.5mm] (sp) edge[bend right=10] node[above] {(1) request} (user);
\draw[->, line width=0.5mm] (user) edge[bend right=10] node[above] {(2) identify} (ap);
\draw[->, line width=0.5mm] (ap) edge[bend right=10] node[above] {(3) identify} (ip);
\draw[->, line width=0.5mm] (ip) edge[bend right=10] node[below] {(4) CP($u$)} (ap);
\draw[->, line width=0.5mm] (ap) edge[bend right=10] node[below] {(5) AP($u'$)} (user);
\draw[->, line width=0.5mm] (user) edge[bend right=10] node[below] {(6) AP($u'$)} (sp);

\end{tikzpicture}

\caption{\cz{A stylised schematic representation of the \textit{SecureKey
Concierge} (SKC) system.}  \textit{The parties are represented by the symbols
``CP'', ``AP'', ``User'', and ``Service'', and the arrows represent messages
between the parties.  The numbers associated with each arrow show the sequence,
and the symbol following the number represents the contents of a message.
First, the service provider (Service) requests authorisation from the user, who
in turn sends identifying information to the authentication provider (AP) to
share with the certification provider (CP).  If the CP accepts the identifying
information, it sends a signed credential $u$ to the AP, which in turn issues a
new credential $u'$ for consumption by the Service, which can now authorise the
user.}}

\label{f:skc}
\end{center}
\end{figure}

Figure~\ref{f:skc} offers a stylised representation of the SKC architecture, as
interpreted from its online documentation~\cite{skc}.  When a user wants to
access a service, the service provider sends a request to the user
\textit{(1)}\footnote{The numbers in italics correspond to messages in the
figure indicated, in this case Figure~\ref{f:skc}.} for credentials.  The user
then sends encrypted identifying information (for example, bank account login
details) to the authentication provider \textit{(2)}, which in this case is
SKC, which then forwards it to the certification provider \textit{(3)}.  Next,
the certification provider responds affirmatively with a ``meaningless but
unique'' identifier $u$ representing the user, and sends it to the
authentication provider \textit{(4)}.  The authentication provider then
responds by signing its own identifier $u'$ representing the user and sending
the message to the user \textit{(5)}, which in turn passes it along to the
service provider \textit{(6)}.  At this point the service provider can accept
the user's credentials as valid.  The SKC documentation indicates that SKC uses
different, unlinked values of $u'$ for each service provider.

\subsection{A Two-Phase Approach}
\label{ss:dis}

\begin{figure}
\begin{center}
\begin{minipage}[t][][t]{0.34\linewidth}
\begin{center}
\scalebox{0.9}{\begin{tikzpicture}[>=latex, node distance=3cm, font={\sf \small}, auto]\ts

\node (ip) at (-2,4) [csm, draw, ultra thick, text width=4em] {\textbf{CP}};
\node (ap) at (0,0) [csm, draw, ultra thick, text width=4em] {\textbf{AP}};
\node (user) at (2,4) [csm, draw, ultra thick, text width=4em] {\textbf{User}};

\draw[->, line width=0.5mm] (user) edge[bend right=10] node[sloped,above] {(1) identify} (ap);
\draw[->, line width=0.5mm] (ap) edge[bend right=10] node[sloped,above] {(2) identify} (ip);
\draw[->, line width=0.5mm] (ip) edge[bend right=10] node[sloped,below] {(3) CP($u$)} (ap);
\draw[->, line width=0.5mm] (ap) edge[bend right=10] node[sloped,below] {(4) CP($u$)} (user);

\end{tikzpicture}}\\
\textit{{\bf \textit{(5a)}} Setup phase.}
\end{center}
\end{minipage}
\hspace{3em}
\begin{minipage}[t][][t]{0.56\linewidth}
\begin{center}
\scalebox{0.9}{\begin{tikzpicture}[>=latex, node distance=3cm, font={\sf \small}, auto]\ts

\node (ip) at (-2,4) [csm, draw, ultra thick, text width=4em] {\textbf{CP}};
\node (ap) at (0,0) [csm, draw, ultra thick, text width=4em] {\textbf{AP}};
\node (user) at (2,4) [csm, draw, ultra thick, text width=4em] {\textbf{User}};
\node (sp) at (4,0) [csm, draw, ultra thick, text width=4em] {\textbf{Service}};

\draw[->, line width=0.5mm] (sp) edge[bend right=10] node[sloped,above] {(5) request} (user);
\draw[->, line width=0.5mm] (user) edge[bend right=10] node[sloped,above] {(6) request $u$} (ap);
\draw[->, line width=0.5mm] (ap) edge[bend right=10] node[sloped,below] {(7) AP($u'$)} (user);
\draw[->, line width=0.5mm] (user) edge[bend right=10] node[sloped,below] {(8) AP($u'$)} (sp);
\draw[->, line width=0.5mm] (ip) edge[dashed] node[sloped,below,align=center] {
    revoke-one or\\revoke-all
} (ap);

\end{tikzpicture}}\\
\textit{{\bf \textit{(5b)}} Operating phase.}
\end{center}
\end{minipage}

\caption{\cz{A schematic representation of a modified version of the SKC system
with a stateful authentication provider and one-time identifiers for services.}
\textit{The user first establishes credentials in the setup phase (5a).  Then,
when a service provider requests credentials from the user in the operating
phase (5b), the user reaches out to the authentication provider for
verification, which assigns a different identifier $u'$ each time.}}

\label{f:skc2}
\end{center}
\end{figure}
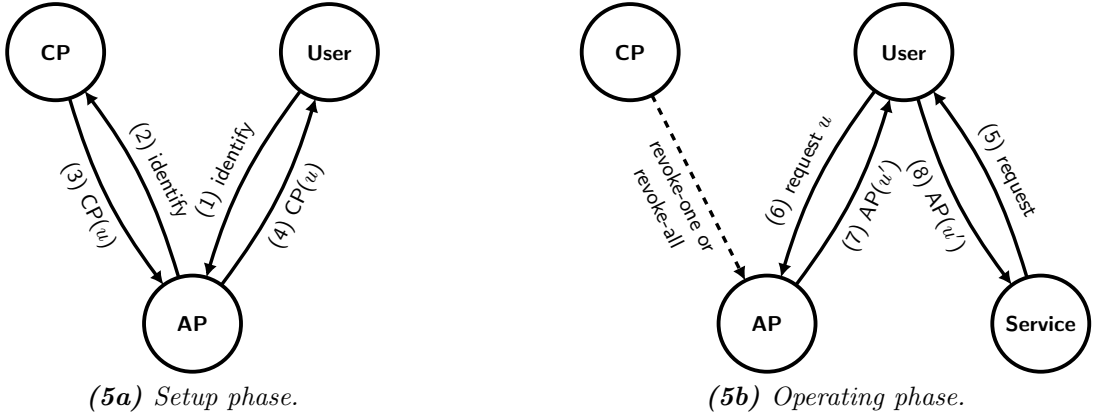

We might consider modifying the SKC architecture so that the user does not need
to log in to the CP each time it requests a service.  To achieve this, we
divide the protocol into two phases, as shown in Figure~\ref{f:skc2}: a
\textit{setup phase} (Figure~\ref{f:skc2}a) in which a user establishes
credentials with an ``certification provider'' (CP) for use with the service,
and an \textit{operating phase} (Figure~\ref{f:skc2}b) in which a user uses the
credentials in an authentication process with a service provider.  So, the
setup phase is done once, and the operating phase is done once per service
request.  In the setup phase, the user first sends authentication credentials,
such as those used to withdraw money from a bank account, to an authentication
provider \textit{(1)}.  The authentication provider then uses the credentials
to authenticate to the certification provider \textit{(2)}, which generates a
unique identifier $u$ that can be used for subsequent interactions with service
providers and sends it to the authentication provider \textit{(3)}, which
forwards it to the user \textit{(4)}.  Then, in the operating phase, a service
provider requests credentials from the user \textit{(5)}, which in turn uses
the previously established unique identifier $u$ to request credentials from
the authentication provider \textit{(6)}.  This means that the user would
implicitly maintain a relationship with the authentication provider, including
a way to log in.  The authentication provider then verifies that the
credentials have not been revoked by the certification provider.  The process
for verifying that the CP credential is still valid may be offline, via a
periodic check, or online, either requiring the AP to reach out to the AP when
it intends to revoke a credential or requiring the AP to send a request to the
CP in real-time.  In the latter case, the AP is looking only for updates on the
set of users who have been through the setup phase, and it does not need to
identify which user has made a request.  Once the AP is satisfied, it sends a
signed certification of its identifier $u'$ to the user \textit{(7)}, which
forwards it to the service provider as before \textit{(8)}.

Unfortunately, even if we can avoid the need for users to log in to the CP
every time they want to use a service, the authentication provider itself
serves as a trusted third party.  Although the SKC architecture may eliminate
the need to trust the existing payment networks, the authentication provider
maintains the mapping between service providers and the certification
providers used by individuals to request services.  It is also implicitly
trusted to manage all of the certification tokens, and there is no way to
ensure that it does not choose them in a way that discloses information to
service providers or certification providers.  In particular, users need to
trust the authentication provider to use identifiers that do not allow service
providers to correlate their activities, and users may also also want to use
different identifiers from time to time to communicate with the \textit{same}
service provider.  As a monopoly platform, it also has the ability to tax or
deny service to certification providers, users, or service providers according
to its own interests, and it serves as a single point of control vulnerable to
exploitation.  For all of these reasons, we maintain that the SKC architecture
remains problematic from a public interest perspective.

\subsection{A User-Oriented Identity Architecture}

In the architecture presented in Section~\ref{ss:dis}, the authentication
provider occupies a position of control.  In networked systems, control points
confer economic advantages on those who occupy them~\cite{trossen2005}, and the
business incentives associated with the opportunity to build platform
businesses around control points have been used to justify their continued
proliferation~\cite{ramakrishnan2017}.

However, control points also expose consumers to risk, not only because the
occupier of the control point may abuse its position but also because the
control point itself creates a vector for attack by third parties.  For both of
these reasons, we seek to prevent an authentication provider from holding too
much information about users.  In particular, we do not want an authentication
provider to maintain a mapping between a user and the particular services that
a user requests, and we do not want a single authentication provider to
establish a monopoly position in which it can dictate the terms by which users
and service providers interact.  For this reason, we put the user, and not the
authentication provider, in the centre of the architecture.

\subsection{Isolation Objectives}
\label{ss:isolation}

For a user to be certain that she is not providing a channel by which
authentication providers can leak her identity or by which service providers
can trace her activity, then she must \textit{isolate} the different
participants in the system.  The constraints allow us to define three
\textit{isolation objectives} as follows:

\begin{enumerate}

\item \textit{Have users generate unlinked identifiers on devices that they own
and trust}.  Unless they generate the identifiers themselves, users have no way
of knowing for sure whether identifiers assigned to them do not contain
personally identifying information.  For users to verify that the identifiers
will not disclose information that might identify them later, they would need
to generate random identifiers using devices and software that they control and
trust.  We suggest that for a user to trust a device, its hardware and software
must be of an open-source, auditable design with auditable provenance.
Although we would not expect that most users would be able to judge the
security properties of the devices they use, open-source communities routinely
provide mechanisms by which users without specialised knowledge can
legitimately conclude, before using new hardware or software, that a diverse
community of experts have considered and approved the security aspects of the
technology.  Examples of such communities include Debian~\cite{debian} for
software and Arduino~\cite{arduino} for hardware, and trustworthy access to
these communities might be offered by local organisations such as libraries or
universities.

\item \textit{Ensure that authentication providers do not learn about the
user's identity or use of services}.  Authentication providers that require
foundational information about a user, or are able to associate different
requests over time with the same user, are in a position to collect information
beyond what is strictly needed for the purpose of the operation.  The role of
the authentication provider is to act as a neutral channel that confers
authority on certification providers, time-shifts requests for credentials,
and separates the certification providers from providers of services.
Performing this function does not require it to collect information about
individual users at any point.

\item \textit{Ensure that information given to service providers is not shared
with authentication providers.}.  The user must be able to credibly trust that
his or her interaction with service providers must remain private.

\end{enumerate}

The communication among the four parties that we propose can be done via
simple, synchronous (e.g. HTTP) protocols that are easily performed by
smartphones and other mobile devices.  The cryptography handling public keys
can be done using standard public-key-based technologies.

\subsection{Repositioning the User to be in the Centre}
\label{ss:uoia}

\begin{figure}
\begin{center}

\begin{minipage}[t][][t]{0.34\linewidth}
\begin{center}
\scalebox{0.9}{\begin{tikzpicture}[>=latex, node distance=3cm, font={\sf \small}, auto]\ts

\node (ip) at (-2,4) [csm, draw, ultra thick, text width=4em] {\textbf{CP}};
\node (ap) at (0,0) [csm, draw, ultra thick, text width=4em] {\textbf{AP}};
\node (user) at (2,4) [csm, draw, ultra thick, text width=4em] {\textbf{User}};

\draw[->, line width=0.5mm] (user) edge[bend right=10] node[sloped,above,align=center] {
    (1) identify,\\${x_1,...,x_n}$
} (ip);
\draw[->, line width=0.5mm] (ip) edge[bend right=10] node[sloped,below,align=center] {
    (2) CP($x_1$),\\...,CP($x_n$)
} (user);
\draw[->, line width=0.5mm] (user) -- node[sloped,below,align=center] {
    (3) CP($x_1$),\\...,CP($x_n$)
} (ap);

\end{tikzpicture}}\\
\textit{{\bf \textit{(6a)}} Setup phase.}
\end{center}
\end{minipage}
\hspace{3em}
\begin{minipage}[t][][t]{0.56\linewidth}
\begin{center}
\scalebox{0.9}{\begin{tikzpicture}[>=latex, node distance=3cm, font={\sf \small}, auto]\ts

\node (ip) at (-2,4) [csm, draw, ultra thick, text width=4em] {\textbf{CP}};
\node (ap) at (0,0) [csm, draw, ultra thick, text width=4em] {\textbf{AP}};
\node (user) at (2,4) [csm, draw, ultra thick, text width=4em] {\textbf{User}};
\node (sp) at (4,0) [csm, draw, ultra thick, text width=4em] {\textbf{Service}};

\draw[->, line width=0.5mm] (sp) edge[bend right=10] node[sloped,above] {(4) request $y$} (user);
\draw[->, line width=0.5mm] (user) edge[bend right=10] node[sloped,above] {
    (5) request $x_i$, [$y$]
} (ap);
\draw[->, line width=0.5mm] (ap) edge[bend right=10] node[sloped,below] {(6) AP([$y$])} (user);
\draw[->, line width=0.5mm] (user) edge[bend right=10] node[sloped,below] {(7) AP($y$)} (sp);
\draw[->, line width=0.5mm] (ip) edge[dashed] node[sloped,below,align=center] {
    revoke-one or\\revoke-all
} (ap);

\end{tikzpicture}}\\
\textit{{\bf \textit{(6b)}} Operating phase.}
\end{center}
\end{minipage}\\

\caption{\cz{A schematic representation of a digital identity system with a
user-oriented approach.} \textit{The new protocol uses user-generated
identifiers and blind signatures to isolate the authentication provider.  The
authentication provider cannot inject identifying information into the
identifiers, nor can it associate the user with the services that she
requests.}}

\label{f:i3}
\end{center}
\end{figure}
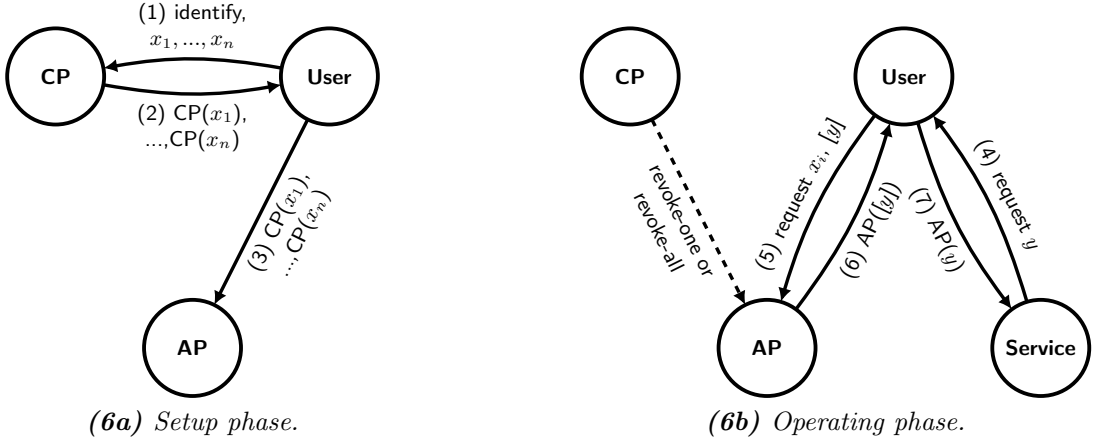

Figure~\ref{f:i3} shows how we can modify the system shown in
Figure~\ref{f:skc2} to achieve the three isolation objectives defined above.
Here, we introduce blind signatures~\cite{chaum82} to allow the user to present
a verified signature without allowing the signer and the relying party to link
the identity of the subject to the subject's legitimate use of a service.

Figure~\ref{f:i3}a depicts the new setup phase. First, on her own trusted
hardware (see Section~\ref{ss:isolation}), the user generates her own set of
identifiers $x_1,...,x_n$ that she intends to use, at most once each, in future
correspondence with the authentication provider.  Generating the identifiers is
not computationally demanding and can be done with an ordinary smartphone.  By
generating her own identifiers, the user has better control that nothing
encoded in the identifiers that might reduce her anonymity.  The user then
sends both its identifying information and the identifiers $x_1,...,x_n$ to the
certification provider \textit{(1)}.  The certification provider then responds
with a set of signatures corresponding to each of the identifiers \textit{(2)}.
The user then sends the set of signatures to the authentication provider for
future use \textit{(3)}.

Figure~\ref{f:i3}b depicts the new operating phase.  First, the service sends a
request to the user along with a new \textit{nonce} (one-time identifier) $y$
corresponding to the request \textit{(4)}.  The user then applies a blinding
function to the nonce $y$, creating a blinded nonce $[y]$.  The user chooses
one of the identifiers $x_i$ that she had generated during the setup phase and
sends that identifier along with the blinded nonce $[y]$ to the authentication
provider \textit{(5)}.  Provided that the signature on $x_i$ has not been
revoked, the authentication provider confirms that it is valid by signing $[y]$
and sending the signature to the user \textit{(6)}.  The user in turn
``unblinds'' the signature on $y$ and sends the unblinded signature to the
service provider \textit{(7)}.  The use of blind signatures ensures that the
authentication provider cannot link what it sees to specific interactions
between the user and the service provider.

\subsection{Architectural Considerations}

To satisfy the constraints listed in Section~\ref{s:constraints}, all three
process steps (identification, authentication, and authorisation) must be
isolated from each other.  Although our proposed architecture introduces
additional interaction and computation, we assert that the complexity of the
proposed architecture is both parsimonious and justified:

\begin{enumerate}

\item If the certification provider were the same as the service provider,
then the user would be subject to direct control and surveillance by that
organisation, violating Constraints 1, 3, and 5.

\item If the authentication provider were the same as the certification
provider, then the user would have no choice but to return to the same
organisation each time it requests a service, violating Constraints 1 and 4.
That organisation would then be positioned to discern patterns in its activity,
violating Constraints 3 and 5.  There would be no separate authentication
provider to face competition for its services as distinct from the
certification services, violating Constraint 6.

\item If the authentication provider were the same as the service provider,
then the service provider would be positioned to compel the user to use a
particular certification provider, violating Constraints 1 and 4.  The service
provider could also impose constraints upon what a certification provider
might reveal about an individual, violating Constraint 3, or how the
certification provider establishes the identity of individuals, violating
Constraint 8.

\item If the user could not generate her own identifiers, then the
certification provider could generate identifiers that reveal information
about the user, violating Constraint 3.

\item If the user were not to use blind signatures to protect the requests from
service providers, then service providers and authentication providers could
compare notes to discern patterns of a user's activity, violating Constraint 5.

\end{enumerate}

The proposed architecture does not achieve its objectives if either the
certification provider or the service provider colludes with the
authentication provider; we assume that effective institutional policy will
complement appropriate technology to ensure that sensitive data are not shared
in a manner that would compromise the interests of the user.

\section{A Decentralised Identity Architecture}
\label{s:dlt}

A significant problem remains with the design described in
Section~\ref{ss:uoia} in that it requires $O(n^2)$ relationships among
authentication providers and certification providers (i.e., with each
authentication provider connected directly to each certification provider that
it considers valid) to be truly decentralised.  Recall that the system relies
critically upon the ability of an certification provider to \textit{revoke}
credentials issued to users, and authentication providers need a way to learn
from the certification provider whether a credential in question has been
revoked.   Online registries such as OCSP~\cite{ocsp}, which are operated by a
certification provider or trusted authority, are a common way to address this
problem, although the need for third-party trust violates Constraint 1.  The
issue associated with requiring each authentication provider to establish its
own judgment of each candidate certification provider is a \textit{business}
problem rather than a technical one.  Hierarchical trust relationships emerge
because relationships are expensive to maintain and introduce risks; all else
being equal, business owners prefer to have fewer of them.  Considered in this
context, concentration and lack of competition among authentication providers
makes sense.  If one or a small number of authentication providers have already
established relationships with a broad set of certification providers, just as
payment networks such as Visa and Mastercard have done with a broad set of
banks, then the cost to a certification provider of a relationship with a new
authentication provider would become a barrier of entry to new authentication
providers.  The market for authentication could fall under the control of a
monopoly or cartel.

\begin{figure}
\begin{center}
\begin{tikzpicture}[>=latex, node distance=3cm, font={\sf \small}, auto]\ts

\node (r1) at (0,0) [circ, draw, dashed, ultra thick, text width=5.2em] {Ledger};
\node (ip) at (2,1.5) [circ, draw, ultra thick, text width=1.3em] {\textbf{CP}};
\node (ap) at (2,-1.5) [circ, draw, ultra thick, text width=1.3em] {\textbf{AP}};
\node (il) at (2,2.3) [circ, align=center, text width=5.2em] {ESTABLISH};
\node (il) at (2,-2.3) [circ, align=center, text width=5.2em] {VERIFY};
\node (r2) at (4,0) [circ, draw, ultra thick, text width=5.2em] {\textbf{User}};
\node (r3) at (8,0) [circ, draw, ultra thick, text width=5.2em] {\textbf{Service}};

\draw[->, line width=0.5mm] (r2) edge[bend right=10] (ip);
\draw[->, line width=0.5mm] (ip) edge[bend right=10] (r1);
\draw[->, line width=0.5mm] (r1) edge[bend right=10] (ap);
\draw[->, line width=0.5mm] (ap) edge[bend right=10] (r2);
\draw[->, line width=0.5mm] (r2) -- node[sloped, above] {ASSERT} (r3);

\end{tikzpicture}

\caption{\cz{A schematic representation of a decentralised identity system with
a distributed ledger.} \textit{The user is not required to interact directly
with the distributed ledger (represented by a dashed circle) and can rely upon
the services offered by certification providers and authentication providers.}}

\label{f:overview-dlt}
\end{center}
\end{figure}
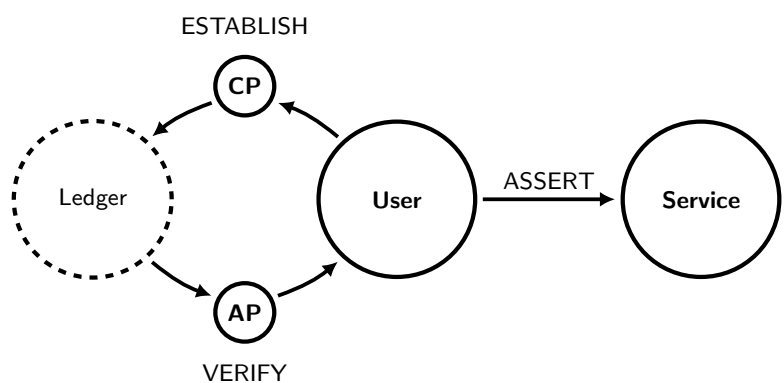

\subsection{Introducing Distributed Ledger Technology}
\label{ss:dlt}

We propose using \textit{distributed ledger technology} (DLT) to allow both
certification providers and authentication providers to proliferate whilst
avoiding industry concentration.  The distributed ledger would serve as a
standard way for certification providers to establish relationships with any
or all of the authentication providers at once, or vice-versa.  The ledger
itself would be a mechanism for distributing signatures and revocations; it
would be shared by participants and not controlled by any single party.
Figure~\ref{f:overview-dlt} shows that users would not interact with the
distributed ledger directly but via their chosen certification providers and
authentication providers.  Additionally, users would not be bound to use any
particular authentication provider when verifying a particular credential and
could even use a different authentication provider each time.  Provided that
the community of participants in the distributed ledger remains sufficiently
diverse, the locus of control would not be concentrated within any particular
group or context, and the market for authentication can remain competitive.

Because the distributed ledger architecture inherently does not require each
new certification provider to establish relationships with all relevant
authentication providers, or vice-versa, it facilitates the entry of new
authentication providers and certification providers, thus allowing the
possibility of decentralisation.

We argue that a distributed ledger is an appropriate technology to maintain the
authoritative record of which credentials have been issued (or revoked) and
which transactions have taken place.  We do not trust any specific third party
to manage the list of official records, and we would need the system to be
robust in the event that a substantial fraction of its constituent parts are
compromised.  The distributed ledger can potentially take many forms, including
but not limited to blockchain, and, although a variety of fault-tolerant
consensus algorithms may be appropriate, we assume that the set of node
operators is well-known, a characteristic that we believe might be needed to
ensure appropriate governance.

If implemented correctly at the system level, the use of a distributed ledger
can ensure that the communication between the certification provider and the
authentication provider is limited to that which is written on the ledger.  If
all blind signatures are done without including any accompanying metadata, and
as long as the individual user does not reveal which blind signature on the
ledger corresponds to the unblinded signature that he or she is presenting to
the authentication provider for approval, then nothing on the ledger will
reveal any information about the individual persons who are the subjects of the
certificates.  We assume that the certification authorities would have a
limited and well-known set of public keys that they would use to sign
credentials, with each key corresponding to the category of individual persons
who have a specific attribute.  The size of the anonymity set for aa
credential, and therefore the effectiveness of the system in protecting the
privacy of an individual user with that credential, depends upon the generality
of the category.  We would encourage certification authorities to assign
categories that are as large as possible.  We would also assume that the
official set of signing keys used by certification providers and authentication
providers is also maintained on the ledger, as doing so would ensure that all
users of the system have the same view of which keys the various certification
providers and authentication providers are using.

\subsection{Achieving Decentralisation with a Distributed Ledger}
\label{ss:fc}

\begin{figure}
\begin{center}
\begin{center}
\scalebox{0.9}{\begin{tikzpicture}[>=latex, node distance=3cm, font={\sf \small}, auto]\ts

\node (ip) at (-2,4) [csm, draw, ultra thick, text width=4em] {\textbf{CP}};
\node (dl) at (-4,0) [csm, draw, ultra thick, dashed, text width=4em] {\textbf{Ledger}};
\node (user) at (2,4) [csm, draw, ultra thick, text width=4em] {\textbf{User}};

\draw[->, line width=0.5mm] (user) -- node[sloped,above,align=center] {
    (1) identify,\\$x_1,...,x_n$
} (ip);
\draw[->, line width=0.5mm] (ip) -- node[sloped,below,align=center] {
    (2) CP($x_1$),\\...,CP($x_n$)
} (dl);

\end{tikzpicture}}\\
\textit{{\bf\textit{(8a)}} Setup phase.}
\end{center}
\vspace{10pt}
\begin{center}
\scalebox{0.9}{\begin{tikzpicture}[>=latex, node distance=3cm, font={\sf \small}, auto]\ts

\node (ip) at (-2,4) [csm, draw, ultra thick, text width=4em] {\textbf{CP}};
\node (dl) at (-4,0) [csm, draw, ultra thick, dashed, text width=4em] {\textbf{Ledger}};
\node (user) at (2,4) [csm, draw, ultra thick, text width=4em] {\textbf{User}};
\node (ap) at (0,0) [csm, draw, ultra thick, text width=4em] {\textbf{AP}};
\node (sp) at (4,0) [csm, draw, ultra thick, text width=4em] {\textbf{Service}};

\draw[->, line width=0.5mm] (sp) edge[bend right=10] node[sloped,above] {(3) request $y$} (user);
\draw[->, line width=0.5mm] (user) edge[bend right=10] node[sloped,above,align=center] {
    (4) prove-owner $x^*_i$,\\request $x_i$, [$y$]
} (ap);
\draw[->, line width=0.5mm] (dl) edge[dashed] node[sloped,above] {updates} (ap);
\draw[->, line width=0.5mm] (ap) edge[bend right=10] node[sloped,below] {(5) AP([$y$])} (user);
\draw[->, line width=0.5mm] (user) edge[bend right=10] node[sloped,below] {(6) AP($y$)} (sp);
\draw[->, line width=0.5mm] (ip) edge[dashed] node[sloped,below,align=center] {
    revoke-one or\\revoke-all
} (dl);

\end{tikzpicture}}\\
\textit{{\bf\textit{(8b)}} ``Online mode'' operating phase.}
\end{center}
\vspace{10pt}
\begin{center}
\scalebox{0.9}{\begin{tikzpicture}[>=latex, node distance=3cm, font={\sf \small}, auto]\ts

\node (ip) at (-2,4) [csm, draw, ultra thick, text width=4em] {\textbf{CP}};
\node (dl) at (-4,0) [csm, draw, ultra thick, dashed, text width=4em] {\textbf{Ledger}};
\node (user) at (2,4) [csm, draw, ultra thick, text width=4em] {\textbf{User}};
\node (ap) at (0,0) [csm, draw, ultra thick, text width=4em] {\textbf{AP}};
\node (sp) at (4,0) [csm, draw, ultra thick, text width=4em] {\textbf{Service}};

\draw[->, line width=0.5mm] (sp) edge[bend right=12] node[sloped,above] {(5) request} (user);
\draw[->, line width=0.5mm] (user) edge[bend right=12] node[sloped,above,align=center] {
    (3) prove-owner $x^*_i$,\\request $x_i$, $u_i$
} (ap);
\draw[->, line width=0.5mm] (dl) edge[dashed] node[sloped,below] {updates} (ap);
\draw[->, line width=0.5mm] (ap) edge[bend right=8] node[sloped,below] {(4) AP$_T$($u_i$)} (user);
\draw[->, line width=0.5mm] (user) edge[bend right=8] node[sloped,below] {(6) AP$_T$($u_i$)} (sp);
\draw[->, line width=0.5mm] (ip) edge[dashed] node[sloped,below,align=center] {
    revoke-one or\\revoke-all
} (dl);

\end{tikzpicture}}\\
\textit{{\bf\textit{(8c)}} ``Offline mode'' operating phase.}
\end{center}

\caption{\cz{A schematic representation of one possible decentralised digital
identity system using a distributed ledger.} \textit{Diagrams (8a) and (8b)
show the setup and operating phases for an initial sketch of our design, which
uses a distributed ledger to promote a scalable marketplace that allows users
to choose certification providers and authentication providers that suit their
needs.  Diagram (8c) shows a variation of the operating phase that can be used
in an offline context, in which the user might not be able to communicate with
an up-to-date authentication provider and the service provider at the same
time.}}

\label{f:dlt}
\end{center}
\end{figure}

Figure~\ref{f:dlt} shows how the modified architecture with the distributed
ledger technology would work.  Figure~\ref{f:dlt}a shows the setup phase.  The
first two messages from the user to the certification provider are similar to
their counterparts in the protocol shown in Figure~\ref{f:i3}a.  However, now
the user also generates $n$ asymmetric key pairs $(x_i,x^*_i)$, where $x_i$ is
the public key and $x^*_i$ is the private key of pair $i$, and it sends each
public key $x_1,...,x_n$ to the certification provider \textit{(1)}.  Then,
rather than sending the signed messages to the authentication provider via the
user, the certification provider then instead writes the signed certificates
directly to the distributed ledger \textit{(2)}.  Importantly, the certificates
would not contain any metadata but only the public key $x_i$ and its bare
signature; eliminating metadata is necessary to ensure that there is no channel
by which a certification provider might inject information that might be later
used to identify a user.  Figure~\ref{f:dlt}b shows the operating phase, which
begins when a service provider asks a user to authenticate and provides some
nonce $y$ as part of the request.

The certification provider can revoke certificates simply by transacting on the
distributed ledger and without interacting with the authentication provider at
all.  Because the user and the authentication provider are no longer assumed to
mutually trust one another, the user must now prove to the authentication
provider that the user holds the private key $x^*_i$ when the user asks the
authentication provider to sign the blinded nonce [$y$] \textit{(4)}.  At this
point we assume that the authentication provider maintains its own copy of the
distributed ledger and has been receiving updates.  The authentication provider
then refers to its copy of the distributed ledger to determine whether a
credential has been revoked, either because the certification provider revoked
a single credential or because the certification provider revoked its own
signing key.  Provided that the credential has not been revoked, the
authentication provider signs the blinded nonce [$y$] \textit{(5)}, which the
user then unblinds and sends to the service provider \textit{(6)}.  The
following messages are carried out as they are done in Figure~\ref{f:i3}b.

We assume that each certification provider and authentication provider has a
distinct signing key for credentials representing each possible policy
attribute, and we further assume that each possible policy attribute admits for
a sufficiently large anonymity set to not identify the user, as described in
Section~\ref{ss:dlt}.  A policy might consist of the union of a set of
attributes, and because users could prevent arbitrary subsets of the attributes
to authentication providers and service providers, we believe that in most
cases it would not be practical to structure policy attributes in such a manner
that one attribute represents a qualification or restriction of another.
Additionally, at a system level, authentication providers and service providers
must not require a set of attributes, either from the same issuer or from
different issuers, whose combination would restrict the anonymity set to an
extent that would potentially reveal the identity of the user.

\subsection{Operating the System Offline}
\label{ss:offline}

The proposed approach can also be adapted to work \textit{offline},
specifically when a user does not have access to an Internet-connected
authentication provider at the time that it requests a service from a service
provider.  This situation applies to two cases: first, in which the
authentication provider has only intermittent access to its distributed ledger
peers (perhaps because the authentication provider has only intermittent access
to the Internet), and second, in which the user does not have access to the
authentication provider (perhaps because the user does not have access to the
Internet) at the time that it requests a service.

In the first case, note that the use of a distributed ledger allows the
authentication provider to avoid the need to send a query in
real-time.\footnote{Not sending the query over the network may also improve the
privacy of the transaction.}  If the authentication provider is disconnected
from the network, then it can use its most recent version of the distributed
ledger to check for revocation.  If the authentication provider is satisfied
that the record is sufficiently recent, then it can sign the record with a key
that is frequently rotated to indicate its timeliness, which we shall denote by
{\sf AP}$_T$.  We presume that {\sf AP}$_T$ is irrevocable but valid for a
limited time only.  If the authentication provider is disconnected from its
distributed ledger peers but still connected to the network with the service
provider, then it can still sign a nonce from the service provider as usual.

In the second case, however, although the user is disconnected from the
network, the service provider still requires an indication of the timeliness of
the authentication provider's signature.  The generalised solution is to adapt
the operating phase of the protocol as illustrated by Figure~\ref{f:dlt}c.
Here, we assume that the user knows in advance that she intends to request a
service at some point in the near future, so she sends the request to the
authentication provider pre-emptively, along with a one-time identifier $u_i$
\textit{(3)}.  Then, the authentication provider verifies the identifier via
the ledger and signs the one-time identifier $u_i$ with the time-specific key
{\sf AP}$_T$ \textit{(4)}.  Later, when the service provider requests
authorisation \textit{(5)}, the user responds with the signed one-time
identifier that it had obtained from the authentication provider \textit{(6)}.
In this protocol, the service provider also has a new responsibility, which is
to keep track of one-time identifiers to ensure that there is no duplication.

\subsection{Achieving Unlinkability with Blinded Credentials}
\label{ss:blind}

\begin{figure}
\begin{center}
\begin{center}
\scalebox{0.9}{\begin{tikzpicture}[>=latex, node distance=3cm, font={\sf \small}, auto]\ts

\node (ip) at (-2,4) [csm, draw, ultra thick, text width=4em] {\textbf{CP}};
\node (dl) at (-4,0) [csm, draw, ultra thick, dashed, text width=4em] {\textbf{Ledger}};
\node (user) at (2,4) [csm, draw, ultra thick, text width=4em] {\textbf{User}};

\draw[->, line width=0.5mm] (user) -- node[sloped,above,align=center] {
    (1) identify,\\$[x_1]$,...,$[x_n]$
} (ip);
\draw[->, line width=0.5mm] (ip) -- node[sloped,above,align=center] {
    (2) at each\\time interval $i$:
} (dl);
\draw[->, line width=0.5mm] (ip) -- node[sloped,below,align=center] {
    CP$_i$([$x_i$])
} (dl);

\end{tikzpicture}}\\
\textit{{\bf\textit{(9a)}} Setup phase.}
\end{center}
\vspace{10pt}
\begin{center}
\scalebox{0.9}{\begin{tikzpicture}[>=latex, node distance=3cm, font={\sf \small}, auto]\ts

\node (ip) at (-2,4) [csm, draw, ultra thick, text width=4em] {\textbf{CP}};
\node (dl) at (-4,0) [csm, draw, ultra thick, dashed, text width=4em] {\textbf{Ledger}};
\node (user) at (2,4) [csm, draw, ultra thick, text width=4em] {\textbf{User}};
\node (ap) at (0,0) [csm, draw, ultra thick, text width=4em] {\textbf{AP}};
\node (sp) at (4,0) [csm, draw, ultra thick, text width=4em] {\textbf{Service}};

\draw[->, line width=0.5mm] (sp) edge[bend right=10] node[sloped,above] {(3) request $y$} (user);
\draw[->, line width=0.5mm] (user) edge[bend right=10] node[sloped,above,align=center] {
    (4) request-certs CP$_i$,\\prove-owner $x^*_i$,\\request CP$_i$($x_i$),[$y$]
} (ap);
\draw[->, line width=0.5mm] (dl) edge[dashed] node[sloped,below] {updates} (ap);
\draw[->, line width=0.5mm] (ap) edge[bend right=10] node[sloped,below] {(5) AP([$y$])} (user);
\draw[->, line width=0.5mm] (user) edge[bend right=10] node[sloped,below] {(6) AP($y$)} (sp);
\draw[->, line width=0.5mm] (ip) edge[dashed] node[sloped,below,align=center] {
    revoke-all
} (dl);

\end{tikzpicture}}\\
\textit{{\bf\textit{(9b)}} ``Online mode'' operating phase.}
\end{center}
\vspace{10pt}
\begin{center}
\scalebox{0.9}{\begin{tikzpicture}[>=latex, node distance=3cm, font={\sf \small}, auto]\ts

\node (ip) at (-2,4) [csm, draw, ultra thick, text width=4em] {\textbf{CP}};
\node (dl) at (-4,0) [csm, draw, ultra thick, dashed, text width=4em] {\textbf{Ledger}};
\node (user) at (2,4) [csm, draw, ultra thick, text width=4em] {\textbf{User}};
\node (ap) at (0,0) [csm, draw, ultra thick, text width=4em] {\textbf{AP}};
\node (sp) at (4,0) [csm, draw, ultra thick, text width=4em] {\textbf{Service}};

\draw[->, line width=0.5mm] (sp) edge[bend right=12] node[sloped,above] {(5) request} (user);
\draw[->, line width=0.5mm] (user) edge[bend right=12] node[sloped,above,align=center] {
    (3) request-certs CP$_i$,\\prove-owner $x^*_i$,\\request CP$_i$($x_i$), $u_i$
} (ap);
\draw[->, line width=0.5mm] (dl) edge[dashed] node[sloped,below] {updates} (ap);
\draw[->, line width=0.5mm] (ap) edge[bend right=8] node[sloped,below] {(4) AP$_T$($u_i$)} (user);
\draw[->, line width=0.5mm] (user) edge[bend right=8] node[sloped,below] {(6) AP$_T$($u_i$)} (sp);
\draw[->, line width=0.5mm] (ip) edge[dashed] node[sloped,below,align=center] {
    revoke-all
} (dl);

\end{tikzpicture}}\\
\textit{{\bf\textit{(9c)}} ``Offline mode'' operating phase.}
\end{center}

\caption{\cz{A schematic representation of a metadata-resistant decentralised
identity architecture.} \textit{This version of the design represents our
recommendation for a generalised identity architecture.  By writing only
blinded credentials to the ledger, this version extends the design shown in
Figure~\ref{f:dlt} to resist an attack in which the certification provider can
expose linkages between different credentials associated with the same user.
Diagrams (9a) and (9b) show the setup and operating phases, analogously to the
online example shown in Figure~\ref{f:dlt}; Diagram (9c) shows the
corresponding offline variant.}}

\label{f:blind}
\end{center}
\end{figure}

Unfortunately, the architecture described in Sections~\ref{ss:fc}
and~\ref{ss:offline} has an important weakness as a result of its reliance on
the revocation of user credentials.  Because the credential that an
certification provider posts to the ledger is specifically identified by the
user at the time that the user asks the authentication provider to verify the
credential, the certification provider may collude with individual
authentication providers to determine when a user makes such requests.  Even
within the context of the protocol, an unscrupulous (or compromised, or
coerced) certification provider may post revocation messages for all of the
credentials associated with a particular user, hence linking them to each
other.

For this reason, we recommend modifying to the protocol to prevent this attack
by using blinded credentials to improve its metadata-resistance.
Figure~\ref{f:blind} shows how this would work.  Rather than sending the public
keys $x_i$ directly to the certification provider, the user sends blinded
public keys [$x_i$], one for each of a series of specific, agreed-upon time
intervals \textit{(1)}.  which in turn would be signed by the certification
provider using a blind signature scheme that does not allow revocation
\textit{(1)}.  The certification provider would not sign all of the public keys
and publish the certificates to the ledger immediately; instead, it would sign
them and post the certificates to the ledger at the start of each time interval
$i$, in each instance signing the user keys with a key of its own specific to
that time interval, {\sf CP}$_i$ \textit{(2)}.  If a user expects to make
multiple transactions per time interval and desires those transactions to
remain unlinked from each other, the user may send multiple keys for each
interval.

When the time comes for the user to request a service, the user must
demonstrate that it is the owner of the private key corresponding to a
(blinded) public key that had been signed by the certification provider.  So
the user must first obtain the set of all certificates signed by {\sf CP}$_i$,
which it can obtain from the authentication provider via a specific request,
{\sf request-certs}.  Then it can find the blind signature on [$x_i$] from the
list and unblind the signature to reveal {\sf CP}$_i(x_i)$.  It can then send
this signature to the authentication provider along with its proof of ownership
of $x^*_i$ as before.

This version of the protocol is the one that we recommend for most purposes.
Although the {\sf request-certs} exchange might require the user to download a
potentially large number of certificates, such a requirement would hopefully
indicate a large anonymity set.  In addition, there may be ways to mitigate the
burden associated by the volume of certificates loaded by the client.  For
example, we might assume that the service provider offers a high-bandwidth
internet connection that allows the user to request the certificates
anonymously from an authentication provider.  Alternatively, we might consider
having the certification provider subdivide the anonymity set into smaller sets
using multiple well-known public keys rather than a single {\sf CP}$_i$, or we
might consider allowing an interactive protocol between the user and the
authentication provider in which the user voluntarily opts to reduce her
anonymity set, for example by specifying a small fraction of the bits in
[$x_i$] as a way to request only a subset of the certificates.

\subsection{Adapting the Design for Spending Tokens}

\begin{figure}
\begin{center}
\begin{center}
\scalebox{0.9}{\begin{tikzpicture}[>=latex, node distance=3cm, font={\sf \small}, auto]\ts

\node (ip) at (-2,4) [csm, draw, ultra thick, text width=4em] {\textbf{CP}};
\node (dl) at (-4,0) [csm, draw, ultra thick, dashed, text width=4em] {\textbf{Ledger}};
\node (user) at (2,4) [csm, draw, ultra thick, text width=4em] {\textbf{User}};
\node (sp) at (4,0) [csm, draw, ultra thick, text width=4em] {\textbf{Service}};

\draw[->, line width=0.5mm] (sp) -- node[sloped,below,align=center] {
    (1) service details
} (user);
\draw[->, line width=0.5mm] (user) edge[bend right=15] node[sloped,above,align=center] {
    (2) identify,\\${[x_1],...,[x_n]}$
} (ip);
\draw[->, line width=0.5mm] (ip) -- node[sloped,below,align=center] {
    (3) $n$ tokens
} (dl);
\draw[->, line width=0.5mm] (ip) edge[bend right=15] node[sloped,below,align=center] {
    (4) CP([$x_1$]),\\...,CP([$x_n$])
} (user);

\end{tikzpicture}}\\
\textit{{\bf\textit{(10a)}} Setup phase.}
\end{center}
\vspace{10pt}
\begin{center}
\scalebox{0.9}{\begin{tikzpicture}[>=latex, node distance=3cm, font={\sf \small}, auto]\ts

\node (ip) at (-2,4) [csm, draw, ultra thick, text width=4em] {\textbf{CP}};
\node (dl) at (-4,0) [csm, draw, ultra thick, dashed, text width=4em] {\textbf{Ledger}};
\node (user) at (2,4) [csm, draw, ultra thick, text width=4em] {\textbf{User}};
\node (ap) at (0,0) [csm, draw, ultra thick, text width=4em] {\textbf{AP}};
\node (sp) at (4,0) [csm, draw, ultra thick, text width=4em] {\textbf{Service}};

\draw[->, line width=0.5mm] (sp) edge[bend right=15] node[sloped,above] {(5) request} (user);
\draw[->, line width=0.5mm] (user) edge[bend right=15] node[sloped,above] {
    (6) CP($x_i$),$x_i$(Service)
} (ap);
\draw[->, line width=0.5mm] (ap) edge[bend right=10] node[sloped,above,align=center] {
    (7) CP($x_i$),\\$x_i$(Service)
} (dl);
\draw[->, line width=0.5mm] (dl) edge[bend right=10] node[sloped,below,align=center] {
    (8) receipt
} (ap);
\draw[->, line width=0.5mm] (ap) edge[bend right=5] node[sloped,below] {(9) AP(receipt)} (user);
\draw[->, line width=0.5mm] (user) edge[bend right=5] node[sloped,below] {(10) AP(receipt)} (sp);
\draw[->, line width=0.5mm] (ip) edge[dashed] node[sloped,below] {revoke-all} (dl);

\end{tikzpicture}}\\
\textit{{\bf\textit{(10b)}} ``Online mode'' operating phase.}
\end{center}
\vspace{10pt}
\begin{center}
\scalebox{0.9}{\begin{tikzpicture}[>=latex, node distance=3cm, font={\sf \small}, auto]\ts

\node (ip) at (-2,4) [csm, draw, ultra thick, text width=4em] {\textbf{CP}};
\node (dl) at (-4,0) [csm, draw, ultra thick, dashed, text width=4em] {\textbf{Ledger}};
\node (user) at (2,4) [csm, draw, ultra thick, text width=4em] {\textbf{User}};
\node (ap) at (0,0) [csm, draw, ultra thick, text width=4em] {\textbf{AP}};
\node (sp) at (4,0) [csm, draw, ultra thick, text width=4em] {\textbf{Service}};

\draw[->, line width=0.5mm] (sp) edge[bend right=10] node[sloped,above] {(9) request} (user);
\draw[->, line width=0.5mm] (user) edge[bend right=10] node[sloped,above] {
    (5) CP($x_i$),$x_i$(Service)
} (ap);
\draw[->, line width=0.5mm] (ap) edge[bend right=10] node[sloped,above,align=center] {
    (6) CP($x_i$),\\$x_i$(Service)
} (dl);
\draw[->, line width=0.5mm] (dl) edge[bend right=10] node[sloped,below,align=center] {
    (7) receipt
} (ap);
\draw[->, line width=0.5mm] (ap) edge[bend right=10] node[sloped,below] {(8) object} (user);
\draw[->, line width=0.5mm] (user) edge[bend right=10] node[sloped,below] {(10) object} (sp);
\draw[->, line width=0.5mm] (ip) edge[dashed] node[sloped,below] {revoke-all} (dl);

\end{tikzpicture}}\\
\textit{{\bf\textit{(10c)}} ``Offline mode'' operating phase.}
\end{center}

\caption{\cz{A schematic representation of a decentralised identity
architecture for exchanging tokens.} \textit{The protocol represented in this
figure can be used to allow users to pay for services privately.  Diagrams (10a)
and (10b) show the setup and operating phases, analogously to the online example
shown in Figures~\ref{f:dlt} and~\ref{f:blind}; Diagram (10c) shows the
corresponding offline variant.}}

\label{f:dlt-tokens}
\end{center}
\end{figure}

The architecture defined in Section~\ref{ss:blind} can also be adapted to allow
users to spend tokens on a one-time basis.  This option may be of particular
interest for social and humanitarian services, in which tokens to be used to
purchase food, medicine, or essential services may be issued by a government
authority or aid organisation to a community at large.  In such cases, the
human rights constraints are particularly important.  Figure~\ref{f:dlt-tokens}
shows how a certification provider might work with an issuer of tokens to
issue spendable tokens to users, who may in turn represent themselves or
organisations.

Figure~\ref{f:dlt-tokens}a shows the setup phase.  We assume that the service
provider first tells the user that it will accept tokens issued by the
certification provider \textit{(1)}.  The user then sends a set of $n$
newly-generated public keys, along with any required identity or credential
information, to the certification provider \textit{(2)}.  We assume that the
tokens are intended to be fungible, so the certification provider issues $n$
new, fungible tokens on the ledger \textit{(3)}.  We need not specify the
details of how the second message works in practice; depending upon the trust
model for the ledger it may be as simple as signing a statement incrementing
the value associated with the certification provider's account, or it may be a
request to move tokens explicitly from one account to another.  Then, the
certification provider signs a set of $n$ messages, each containing one of the
blinded public keys, and sends them to the user \textit{(4)}.  The messages
will function as promissory notes, redeemable by the user that generated the
keys, for control over the tokens.

Figure~\ref{f:dlt-tokens}b shows the operating phase.  When a service provider
requests a token \textit{(5)}, the user sends a message to an authentication
provider demonstrating both that it has the right to control the token issued
by the certification provider and that it wishes to sign the token over to the
service provider \textit{(6)}.  The authentication provider, who never learns
identifying information about the user, lodges a transaction on the ledger that
assigns the rights associated with the token to the service provider
\textit{(7)}, generating a receipt \textit{(8)}.  Once the transaction is
complete, the authentication provider shares a receipt with the user
\textit{(9)}, which the user may then share with the service provider
\textit{(10)}, who may now accept that the payment as complete.

Like the ``main'' architecture described in Section~\ref{ss:blind}, the
``token'' architecture can also be configured to work in an offline context by
modifying the operating phase.  Figure~\ref{f:dlt-tokens}c shows how this would
work.  The user requests from the authentication provider one or more physical
``objects'', which may take the form of non-transferable electronic receipts
or physical tokens, that can be redeemed for services from the service provider
\textit{(5)}.  The authentication provider sends the objects to the user
\textit{(8)}, who then redeems them with the service provider in a future
interaction \textit{(9,10)}.

\section{Governance Considerations}
\label{s:governance}

An important challenge that remains with the distributed ledger system
described in Section~\ref{s:dlt} is the management of the organisations that
participate in the consensus mechanism of the distributed ledger.  We believe
that this will require the careful coordination of local businesses and
cooperatives to ensure that the system itself does not impose any
non-consensual trust relationships (Constraint 4), that no single market
participant would gain dominance (Constraint 6), and that participating
businesses and cooperatives will be able to continue to establish their own
business practices and trust relationships on their own terms (Constraint 7),
even while consenting to the decisions of the community of organisations
participating in the shared ledger.  We believe that our approach will be
enhanced by the establishment of a multi-stakeholder process to develop the
protocols by which the various parties can interact, including but not limited
to those needed to participate in the distributed ledger, and ultimately
facilitate a multiplicity of different implementations of the technology needed
to participate.  Industry groups and regulators will still need to address the
important questions of determining the rules and participants.  We surmise that
various organisations, ranging from consulting firms to aid organisations,
would be positioned to offer guidance to community participants in the network,
without imposing new constraints.

\subsection{Open Standards}

The case for a common, industry-wide platform for exchanging critical data via
a distributed ledger is strong.  Analogous mechanisms have been successfully
deployed in the financial industry.  Established mechanisms take the form of
centrally-managed cooperative platforms such as SWIFT~\cite{swift}, which
securely carries messages on behalf of financial markets participants, while
others take the form of consensus-based industry standards, such as the
Electronic Data Interchange (EDI) standards promulgated by X12~\cite{x12} and
EDIFACT~\cite{edifact}.  Distributed ledgers such as Ripple~\cite{ripple} and
Hyperledger~\cite{hyperledger} have been proposed to complement or replace the
existing mechanisms.

For the digital identity infrastructure, we suggest that the most appropriate
application for the distributed ledger system described in Section~\ref{s:dlt}
would be a technical standard for business transactions promulgated by a
self-regulatory organisation working concordantly with government regulators.
A prime example from the financial industry is \textit{best execution},
exemplified by Regulation NMS~\cite{regnms}, which led to the dismantling of a
structural monopoly in electronic equities trading in the United
States.\footnote{See also MiFID, its European Union counterpart~\cite{mifid}.}
Although the US Securities and Exchange Commission had the authority to compel
exchanges to participate in a national market system since 1975, it was not
until thirty years later that the SEC moved to explicitly address the monopoly
enjoyed by the New York Stock Exchange (NYSE).  The Order Protection Rule
imposed by the 2005 regulation (Rule 611) ``requir[ed] market participants to
honor the best prices displayed in the national market system by automated
trading centers, thus establishing a critical linkage
framework''~\cite{regnms-timeline}.  The monopoly was broken, NYSE member firms
became less profitable, and NYSE was ultimately bought by Intercontinental
Exchange in 2013~\cite{ice}.

We believe that distributed ledgers offer a useful mechanism by which
self-regulatory organisations satisfy regulations precisely intended to prevent
the emergence of control points, market concentration, or systems whose design
reflects a conflict of interest between their operators and their users.

\subsection{No Master Keys, No Early-Binding}

An important expectation implicit to the design of our system is that users can
establish and use as many identities as they want, without restriction.  This
means not only that a user can choose which credentials to show to relying
parties, but also that a user would not be expected to bind the credentials to
each other in any way prior to their use.  Such a binding would violate
Constraint 5 from Section~\ref{s:constraints}.  In particular, given two
credentials, there should be no way to know or prove that they were issued to
the same individual or device.  This property is not shared by some schemes for
non-transferable anonymous credentials that encourage users to bind together
credentials to each other via a master key or similar
mechanism~\cite{camenisch2001} as described in Section~\ref{ss:da}.

If it were possible to prove that two or more credentials were associated with
the same identity, then an individual could be forced to associate a set of
credentials with each other inextricably, and even if an individual might be
given an option to reveal only a subset of his or her credentials to a service
provider at any given time, the possibility remains that an individual might be
compelled to reveal the linkages between two or more credentials.  For example,
the device that an individual uses might be compromised, revealing the master
key directly, which would be problematic if the same master key were used for
many or all of the individual's credentials.  Alternatively, the individual
might be coerced to prove that the same master key had been associated with two
or more credentials.

The system we describe explicitly does not seek to rely upon the \textit{ex
ante} binding together of credentials to achieve non-transferability or for any
other purpose.  We suggest that the specific desiderata and requirements for
non-transferability might vary across use cases and can be addressed
accordingly.  Exogenous approaches to achieve non-transferability might have
authentication providers require users to store credentials using trusted
escrow services or physical hardware with strong counterfeiting resistance such
as two-factor authentication devices.  Endogenous approaches might have
authentication providers record the unblinded signatures on the ledger once
they are presented for inspection, such that multiple uses of the same
credential become explicitly bound to each other \textit{ex post} or are
disallowed entirely.  Recall that the system assumes that credentials are used
once only and that certification providers would generate new credentials for
individuals in limited batches, for example at a certain rate over time.

\section{Use Cases}

\begin{table}
\begin{center}
\begin{tikzpicture}[>=latex, node distance=3cm, font={\sf \small}, auto]\ts
\node (r1) at (0,0) [block] {library access\\\vspace{5pt}\\programme eligibility};
\node (r2) at (0,3.52) [block] {membership\\programmes};
\node (r3) at (3.52,0) [block] {public transportation};
\node (r4) at (3.52,3.52) [block] {mobile communication\\services};
\node (a1) at (0,5.62) [noshape, text width=8em] {assert entitlements};
\node (a2) at (-2.1,0) [noshape, rotate=90] {public sector};
\node (a3) at (3.52,5.6) [noshape] {spend tokens};
\node (a4) at (-2.1,3.52) [noshape, rotate=90, text width=8em] {private sector};

\end{tikzpicture}

\caption{\textit{A categorisation matrix for use cases.}  We divide the
universe of use cases into four categories based upon whether the purpose is to
assert entitlements or to spend tokens and upon whether the services in question are
operated by the public sector or the private sector, and we include some examples.}

\label{t:use-cases}
\end{center}
\end{table}

We anticipate that there might be many potential use cases for a decentralised
digital identity infrastructure that affords users the ability to manage the
linkages among their credentials.  Table~\ref{t:use-cases} offers one view of
how the use cases might be divided into four categories on the basis of whether
the purpose is to assert entitlements or spend tokens and upon whether the
services in question are operated by the public sector or the private sector.
Use cases that involve asserting entitlements might include asserting
membership of a club for the purpose of gaining access to facilities, accessing
restricted resources, or demonstrating eligibility for a discount, perhaps on
the basis of age, disability, or financial hardship, at the point of sale.  Use
cases that involve spending tokens can potentially be disruptive, particularly
in areas that generate personally identifiable information.  We imagine that a
decentralised digital identity infrastructure that achieves the privacy
requirements would be deployed incrementally, whether general purpose or not.
We suggest the following three use cases might be particularly appropriate
because of their everyday nature, and might be a fine place to start:

\begin{enumerate}

\item {\bf\textit{Access to libraries.}}  Public libraries are particularly
sensitive to risks associated with surveillance~\cite{zimmer2018}.  The
resources of a public library are the property of its constituency, and the
users have a particular set of entitlements that have specific limitations in
terms of time and quantity.  Entitlement levels could be managed by having the
issuer use a different signing key for each entitlement.  User limits could be
enforced in several ways.  One method involves requiring a user to make a
security deposit that is released at the time that a resource has been returned
and determined to be suitable for recirculation.  Another method involves
requiring the library to check the ledger to verify that a one-time credential
has not already been used as a precondition for providing the resource and
requiring the user purchase the right to a one-time credential that can only be
re-issued upon return of the resource.

\item {\bf\textit{Public transportation.}}  It is possible to learn the habits,
activities, and relationships of individuals by monitoring their trips in an
urban transportation system, and the need for a system-level solution has been
recognised~\cite{heydt2006}.  Tokens for public transportation (for example,
pay-as-you-go or monthly bus tickets) could be purchased with cash in one
instance and then spent over time with each trip unlinkable to each of the
others.  This can be achieved by having an issuer produce a set of one-time use
blinded tokens in exchange for cash and having a user produce one token for
each subsequent trip.  Time-limited services such as monthly travel passes
could be issued in bulk, including a signature with a fixed expiration date
providing a sufficiently large anonymity set.  An issuer could also create
tokens that might be used multiple times, subject to the proviso that trips for
which the same token is used could be linked.

\item {\bf\textit{Wireless data service plans.}} Currently, many mobile devices
such as phones contain unique identifiers that are linked to accounts by
service providers and can be identified when devices connect to cellular
towers~\cite{imei}.  However, it is not actually technically necessary for
service providers to know the particular towers to which a specific customer is
connecting.  For the data service business to be tenable, we suggest that what
service providers really need is a way to know that their customers have paid.
Mobile phone service subscribers could have their devices present blinded
tokens, obtained from issuers following purchases at sales offices or via
subscription plans, to cellular towers without revealing their specific
identities, thus allowing them to avoid tracking over extended periods of time.
Tokens might be valid for a limited amount of time such as an hour, and a
customer would present a token to receive service for a limited time.  System
design considerations would presumably include tradeoffs between the degree of
privacy and the efficiency of mobile handoff between towers or time periods.

\end{enumerate}

We do not anticipate or claim that our system will be suitable for all purposes
for which an individual might be required to present electronic credentials.
We would imagine that obtaining security clearances or performing certain
duties associated with public office might explicitly require unitary identity.
Certain activities related to national security undertaken by ordinary persons,
such as crossing international borders, might also fall into this category,
although we argue that such use cases must be narrowly circumscribed to offer
limited surveillance value through record linkage.  In particular, linking any
strongly non-transferable identifiers or credentials to the identities that
individuals use for routine activities (such as social media, for example, or
the use cases described above) would specifically compromise the privacy rights
of their subjects.  Other application domains, such as those involving public
health or access to medical records, present specific complications that might
require a different design.  Certain financial activities would require
interacting with regulated financial intermediaries who are subject to AML and
KYC regulations, as mentioned in Section~\ref{ss:skc}.  For this reason,
achieving privacy for financial transactions might require a different approach
that operates with existing financial regulations~\cite{goodell2019}.

\section{Conclusions and Future Work}

We argue that the ability of individuals to create and maintain multiple
unrelated identities is a fundamental, inalienable human right.  For the
digital economy to function while supporting this human right, individuals must
be able to control and limit what others can learn about them over the course
of their many interactions with services, including government and
institutional services.

We have introduced a framework for an open digital identity architecture that
promotes the implementation of identity architectures that satisfy constraints
that we consider essential to the protection of human rights, and we believe
that a combination of strong technology and thoughtful policy will be necessary
to promote and ensure the implementation, deployment, and use of technology
that satisfies them.  We have elaborated eight requirements for technology
infrastructure and demonstrated that they can be achieved by means of a
decentralised architecture.  Our framework does not seek strong
non-transferability via an early-binding approach, and we argue that
distributed ledgers can be used not only to achieve the privacy objectives but
also to deliver an alternative to strong non-transferability.  We have
identified challenges associated with scalability and governance, and we have
also demonstrated how tokens can be spent via such a system as well as how the
system might be used in an offline context.

Future work may include formal analysis of the information security properties
of a system designed according to this framework, as well as the development of
a proof-of-concept implementation and a corresponding evaluation of the various
implementation tradeoffs relevant to different use cases.  We suggest that
different use cases would entail significantly different design choices.

The specific mechanism for fostering a community of participating organisations
will depend upon the relationship between those organisations and the group
that ultimately assumes the role of ensuring that the system does not impose
non-consensual trust relationships on its users.  It must be noted that any
system that puts control in the hands of end-users carries the burden of
education, both for the well-functioning of the system as well as for
safeguarding its role in protecting the public interest.  Future research,
therefore, must include case studies of how similar systems have been
developed, deployed, and maintained over time, in a variety of different social
and business contexts.

\section{Acknowledgements}

We thank Valerie Khan, Edgar Whitley, Paul Makin, and Oscar King for their
thoughtful insights.  Geoff Goodell is also an associate of the Centre for
Technology and Global Affairs at the University of Oxford.  We acknowledge the
Engineering and Physical Sciences Research Council (EPSRC) for the BARAC
project (EP/P031730/1) and the European Commission for the FinTech project
(H2020-ICT-2018-2 825215).  Tomaso Aste acknowledges the Economic and Social
Research Council (ESRC) for funding the Systemic Risk Centre (ES/K0 02309/1).
This manuscript has been released as a Pre-Print at
\url{http://export.arxiv.org/pdf/1902.08769}.


\begin{thebibliography}{1}\raggedright




\bibitem{id4d}{
    The World Bank.
    ``Identification for Development.''
    [online]
    \url{https://id4d.worldbank.org/}
    [retrieved 2019-10-09]
}
\bibitem{rohingya}{
    Rohingya Project: Unlocking Potential.
    [online]
    \url{http://rohingyaproject.com/}
    [retrieved 2019-06-05]
}
\bibitem{evernym}{
    Evernym: The Self-Sovereign Identity Company.
    [online]
    \url{https://www.evernym.com/}
    [retrieved 2019-10-09]
}
\bibitem{id2020}{
    ID2020: Digital Identity Alliance.
    [online]
    \url{https://id2020.org/}
    [retrieved 2019-10-09]
}
\bibitem{5rights}{
    5Rights Foundation.
    [online]
    \url{https://5rightsfoundation.com/}
    [retrieved 2019-10-09]
}
\bibitem{pandya2019}{
    Pandya, J.
    ``Hacking Our Identity: The Emerging Threats From Biometric Technology.''
    \textit{Forbes},
    2019-03-09.
    [online]
    \url{https://www.forbes.com/sites/cognitiveworld/2019/03/09/hacking-our-identity-the-emerging-threats-from-biometric-technology/}
    [retrieved 2019-10-09]
}
\bibitem{camenisch2001}{
    Camenisch, J and Lysyanskaya, A.
    ``An Efficient System for Non-transferable Anonymous Credentials with Optional Anonymity Revocation.''
    Proceedings of the International Conference on the Theory and Applications
    of Cryptographic Techniques (EUROCRYPT 2001: Advances in Cryptology),
    2001-04-15,
    pp. 93--118.
    [online]
    \url{https://eprint.iacr.org/2001/019.pdf}
    [retrieved 2019-07-25]
}
\bibitem{abelson1997}{
    H. Abelson, R. Anderson, S. Bellovin, J. Benaloh, M. Blaze, W. Diffie, J. Gilmore, P. Neumann, R. Rivest, J. Schiller, and B. Schneier.
    ``The Risks of Key Recovery, Key Escrow, and Trusted Third-Party Encryption.''
    \texttt{doi:10.7916/D8GM8F2W},
    1997-05-27.
    [online]
    \url{https://academiccommons.columbia.edu/doi/10.7916/D8R2176H/download}
    [retrieved 2019-03-11]
}
\bibitem{abelson2015}{
    H. Abelson, R. Anderson, S. Bellovin, J. Benaloh, M. Blaze, W. Diffie, J. Gilmore, M. Green, S. Landau, P. Neumann, R. Rivest, J. Schiller, B. Schneier, M. Specter, and D. Weitzner.
    ``Keys under doormats: mandating insecurity by requiring government access to all data and communications.''
    \textit{Journal of Cybersecurity} \textbf{1}(1),
    pp. 69--79,
    \texttt{doi:10.1093/cybsec/tyv009},
    2015-11-17.
    [online]
    \url{https://academiccommons.columbia.edu/doi/10.7916/D82N5D59/download}
}
\bibitem{benaloh2018}{
    J. Benaloh.
    ``What if Responsible Encryption Back-Doors Were Possible?''
    Lawfare Blog,
    2018-11-29.
    [online]
    \url{https://www.lawfareblog.com/what-if-responsible-encryption-back-doors-were-possible}
    [retrieved 2018-12-11]
}
\bibitem{kuner2017}{
    Kuner, C and Marelli, M.
    \textit{Handbook on Data Protection in Humanitarian Action.}
    International Committee of the Red Cross,
    Geneva, Switzerland,
    2017-08-23.
    [online]
    \url{https://www.icrc.org/en/publication/handbook-data-protection-humanitarian-action}
    [retrieved 2019-10-01]
}
\bibitem{stevens2018}{
    Stevens, L.
    ``Self-Sovereign Identity Systems for Humanitarian Interventions.''
    Working Paper.
    [online]
    \url{https://pdfs.semanticscholar.org/f821/4975160857f1f020ff8dbc2db65f88fcac03.pdf}
    [retrieved 2019-10-01]
}
\bibitem{privacy-identity}{
    Privacy International.
    ``Identity.''
    [online]
    \url{https://privacyinternational.org/topics/identity}
    [retrieved 2019-10-01]
}
\bibitem{fatf-recommendations}{
    Financial Action Task Force (FATF).
    \textit{The FATF Recommendations.}
    Updated February 2018.
    [online]
    \url{https://web.archive.org/web/20190718073849/http://www.fatf-gafi.org/media/fatf/documents/recommendations/pdfs/FATF%2520Recommendations%25202012.pdf}
    [retrieved 2018-09-16]
}
\bibitem{snoopers}{
    Parliament of the United Kingdom.
    ``Investigatory Powers Bill: Committee Stage Report.''
    House of Commons Library (Commons Briefing papers CBP-7578),
    2016-06-02.
    [online]
    \url{https://researchbriefings.parliament.uk/ResearchBriefing/Summary/CBP-7578}
    [retrieved 2019-10-09]
}
\bibitem{australia2018}{
    Parliament of Australia.
    ``Telecommunications and Other Legislation Amendment (Assistance and Access) Bill 2018.''
    [online]
    \url{https://www.aph.gov.au/Parliamentary_Business/Bills_Legislation/Bills_Search_Results/Result?bId=r6195}
    [retrieved 2019-10-01]
}
\bibitem{zuboff2015}{
    S. Zuboff.
    ``Big Other: surveillance capitalism and the prospects of an information civilization.''
    \textit{Journal of Information Technology}
    \textbf{30},
    2015.
    pp. 75--89.
    [online]
    \url{http://www.palgrave-journals.com/jit/journal/v30/n1/pdf/jit20155a.pdf}
    [retrieved 2018-09-17]
}
\bibitem{waldman}{
    P. Waldman, L. Chapman, and J. Robertson.
    ``Palantir Knows Everything About You.''
    Bloomberg,
    2018-04-19.
    [online]
    \url{https://www.bloomberg.com/features/2018-palantir-peter-thiel/}
    [Retrieved 2018-04-19]
}
\bibitem{zooko}{
    Wilcox-O'Hearn, Z.
    ``Names: Decentralized, Secure, Human-Meaningful: Choose Two.''
    [online]
    \url{https://web.archive.org/web/20011020191610/http://zooko.com/distnames.html}
    [retrieved 2018-04-21]
}
\bibitem{petnames}{
    Stiegler, M.
    ``An Introduction to Petname Systems.''
    February 2005.
    [online]
    \url{http://www.skyhunter.com/marcs/petnames/IntroPetNames.html}
    [retrieved 2018-05-11]
}
\bibitem{douceur}{
    Douceur, J.
    ``The Sybil Attack.''
    \textit{IPTPS '01 Revised Papers from the First International Workshop on Peer-to-Peer Systems},
    2002-03-07,
    pp. 251--260.
    [online]
    \url{https://www.freehaven.net/anonbib/cache/sybil.pdf}
    [retrieved 2018-05-11]
}
\bibitem{vanderburg}{
    Vanderburg, E.
    ``A Certified Lack of Confidence: The Threat of Rogue Certificate Authorities.''
    TCDI Blog.
    [online]
    \url{https://www.tcdi.com/the-threat-of-rogue-certificate-authorities/}
    [retrieved 2018-05-11]
}
\bibitem{charette}{
    Charette, R.
    ``DigiNotar Certificate Authority Breach Crashes e-Government in the Netherlands.''
    \textit{IEEE Spectrum},
    2011-09-09.
    [online]
    \url{https://spectrum.ieee.org/riskfactor/telecom/security/diginotar-certificate-authority-breach-crashes-egovernment-in-the-netherlands}
    [retrieved 2018-05-11]
}
\bibitem{stuxnet}{
    Kushner, D.
    ``The Real Story of Stuxnet.''
    \textit{IEEE Spectrum},
    2013-02-26.
    [online]
    \url{https://spectrum.ieee.org/telecom/security/the-real-story-of-stuxnet}
    [retrieved 2018-05-11]
}
\bibitem{eckersley}{
    Eckersley, P.
    Electronic Frontier Foundation Technical Analysis,
    2011-03-23.
    [online]
    \url{https://www.eff.org/deeplinks/2011/03/iranian-hackers-obtain-fraudulent-https}
    [retrieved 2018-05-11]
}
\bibitem{arstechnica3}{
    Goodin, D.
    ``Already on probation, Symantec issues more illegit HTTPS certificates.''
    \textit{Ars Technica},
    2017-01-20.
    [online]
    \url{https://arstechnica.com/information-technology/2017/01/already-on-probation-symantec-issues-more-illegit-https-certificates/}
    [retrieved 2018-05-11]
}
\bibitem{equifax}{
    Equifax Inc.
    ``2017 Cybersecurity Incident \& Important Consumer Information.''
    2018-03-01.
    [online]
    \url{https://www.equifaxsecurity2017.com/}
    [retrieved 2018-05-11]
}
\bibitem{register}{
    Chirgwin, R.
    ``Google publishes list of Certificate Authorities it doesn't trust.''
    \textit{The Register},
    2016-03-23.
    [online]
    \url{https://www.theregister.co.uk/2016/03/23/google_now_publishing_a_list_of_cas_it_doesnt_trust/}
    [retrieved 2018-05-11]
}
\bibitem{slashdot}{
    Slashdot.
    ``School Tricks Pupils Into Installing a Root CA.''
    2014-03-09.
    [online]
    \url{https://news.slashdot.org/story/14/03/09/0225224/school-tricks-pupils-into-installing-a-root-ca}
    [retrieved 2018-05-11]
}
\bibitem{bright}{
    Bright, P.
    ``Gov’t, certificate authorities conspire to spy on SSL users?''
    \textit{Ars Technica},
    2010-03-29.
    [online]
    \url{https://web.archive.org/web/20171004131406/http://files.cloudprivacy.net/ssl-mitm.pdf}
    [retrieved 2018-05-11]
}
\bibitem{arstechnica1}{
    Goodin, D.
    ``Stuxnet-style code signing is more widespread than anyone thought.''
    \textit{Ars Technica}.
    2017-11-03.
    [online]
    \url{https://arstechnica.com/information-technology/2017/11/evasive-code-signed-malware-flourished-before-stuxnet-and-still-does/}
    [retrieved 2018-04-21]
}
\bibitem{arstechnica2}{
    Goodin, D.
    ``Flaw crippling millions of crypto keys is worse than first disclosed.''
    \textit{Ars Technica}.
    2017-11-06.
    [online]
    \url{https://arstechnica.com/information-technology/2017/11/flaw-crippling-millions-of-crypto-keys-is-worse-than-first-disclosed/}
    [retrieved 2018-04-21]
}
\bibitem{engadget1}{
    Moon, M.
    ``Estonia freezes resident ID cards due to security flaw.''
    \textit{Engadget.}
    2017-11-04.
    [online]
    \url{https://www.engadget.com/2017/11/04/estonia-freezes-resident-id-cards-security-flaw/}
    [retrieved 2018-04-21]
}
\bibitem{estonia}{
    Thevoz, P.
    ``Diving into a `Digital Country': e-Estonia.''
    \textit{Medium}.
    2016-09-16.
    [online]
    \url{https://medium.com/@PhilippeThevoz/diving-into-a-digital-country-e-estonia-af561925c95e}
    [retrieved 2018-04-21]
}
\bibitem{aadhaar}{
    Tully, M.
    ``The problem with Aadhaar cards is the way they are being pushed by the State.''
    Hindustan Times,
    2017-12-09.
    [online]
    \url{https://www.hindustantimes.com/analysis/the-problem-with-aadhaar-cards-is-the-way-they-are-being-pushed-by-the-state/story-RTlWUXgF3ck4rsoN1zKXUI.html}
    [retrieved 2018-04-21]
}


\bibitem{wagner2014}{
    Wagner, R.
    ``Identity and Access Management 2020.''
    ISSA Journal \textbf{14}(6),
    June 2014,
    pp. 26--30.
    [online]
    \url{http://www.issa.org/resource/resmgr/JournalPDFs/feature0614.pdf}
    [retrieved 2019-01-11]
}
\bibitem{shuhaimi2012}{
    Shuhaimi, N and Juhana, T.
    ``Security in vehicular ad-hoc network with Identity-Based Cryptography approach: A survey.''
    7th International Conference on Telecommunication Systems, Services, and Applications (TSSA),
    2012-10-30.
    [online]
    \url{http://ieeexplore.ieee.org/abstract/document/6366067/}
    [retrieved 2019-01-11]
}
\bibitem{govukverify}{
    Government Digital Service (UK).
    ``GOV.UK Verify: Guidance.''
    2018-12-14.
    [online]
    \url{https://www.gov.uk/government/publications/introducing-govuk-verify/introducing-govuk-verify}
    [retrieved 2019-02-15]
}
\bibitem{brandao2015}{
    Brandao, L, Christin, N, Danezis, G, and Anonymous.
    ``Toward Mending Two Nation-Scale Brokered Identification Systems.''
    \textit{Proceedings on Privacy Enhancing Technologies} \textbf{2015}(2),
    pp. 135–-155.
    [online]
    \url{http://www0.cs.ucl.ac.uk/staff/G.Danezis/papers/popets15-brokid.pdf}
    [retrieved 2019-01-11]
}
\bibitem{whitley2018}{
    Whitley, E.
    ``Trusted Digital Identity Provision: GOV.UK Verify's Federated Approach.''
    Center for Global Development Policy Paper 131,
    2018-11-07.
    [online]
    \url{https://www.cgdev.org/sites/default/files/Trusted-Digital-ID-Provision-govuk.pdf}
    [retrieved 2019-02-15]
}
\bibitem{ohara2011}{
    O'Hara, K, Whitley, E, and Whittall, P.
    ``Avoiding the Jigsaw Effect: Experiences with Ministry of Justice Reoffending Data.''
    Monograph,
    2011-12-19, modified 2018-06-06.
    [online]
    \url{https://eprints.soton.ac.uk/273072/1/AVOIDING%2520THE%2520JIGSAW%2520EFFECT.pdf}
    [retrieved 2018-07-26]
}
\bibitem{thomas2009}{
    Thomas, T.
    ``Joint Watermarking Scheme for Multiparty Multilevel DRM Architecture.''
    IEEE Transactions on Information Forensics and Security
    \textbf{4}(4),
    2009-09-29,
    pp. 758--767.
    [online]
    \url{http://ieeexplore.ieee.org/abstract/document/5272486/}
    [retrieved 2019-01-11]
}
\bibitem{liu2008}{
    Liu, S, Wei, J, and Li, C.
    ``Method and System for Implementing Authentication on Information Security.''
    United States Patent Application US20080065895A1,
    2008-03-13.
    [online]
    \url{https://patents.google.com/patent/US20080065895A1/en}
    [retrieved 2019-01-11]
}
\bibitem{thackston2018}{
    Thackston, J.
    ``System and method for verifying user identity in a virtual environment.''
    US Patent Grant US10153901B2,
    2018-12-11.
    [online]
    \url{https://patents.google.com/patent/US10153901B2/en}
    [retrieved 2019-01-11]
}


\bibitem{dunphy2018}{
    Dunphy, P and Petitcolas, F.
    ``A First Look at Identity Management Schemes on the Blockchain.''
     arXiv:1801.03294v1 [cs.CR],
    2018-01-10.
    [online]
    \url{https://arxiv.org/abs/1801.03294}
    [retrieved 2019-01-11]
}
\bibitem{shocard}{
    SITA.
    ``Travel Identity of the Future -- White Paper.''
    ShoCard,
    May 2016.
    [online]
    \url{https://shocard.com/wp-content/uploads/2016/11/travel-identity-of-the-future.pdf}
    [retrieved 2019-01-11]
}
\bibitem{everest}{
    Everest.
    ``Everest blockchain software for verified value exchange.''
    [online]
    \url{https://everest.org/}
    [retrieved 2019-10-01]
}
\bibitem{graglia2018}{
    Graglia, M, Mellon, C, and Robustelli, T.
    ``The Nail Finds a Hammer: Self-Sovereign Identity, Design Principles, and Property Rights in the Developing World.''
    \textit{New America},
    2018-10-17.
    [online]
    \url{https://www.newamerica.org/future-property-rights/reports/nail-finds-hammer/exploring-three-platforms-through-the-principles/}
    [retrieved 2019-10-10]
}
\bibitem{sovrin}{
    Tobin, A and Reed, D.
    ``The Inevitable Rise of Self-Sovereign Identity.''
    The Sovrin Foundation,
    2016-09-29.
    [online]
    \url{https://sovrin.org/wp-content/uploads/2017/06/The-Inevitable-Rise-of-Self-Sovereign-Identity.pdf}
    [retrieved 2019-01-11]
}
\bibitem{aitken2018}{
    Aitken, R.
    ``IBM Blockchain Joins Sovrin's 'Decentralized' Digital Identity Network To Stem Fraud.''
    Forbes,
    2018-04-05.
    [online]
    \url{https://www.forbes.com/sites/rogeraitken/2018/04/05/ibm-blockchain-joins-sovrins-decentralized-digital-identity-network-to-stem-fraud/}
    [retrieved 2019-01-11]
}
\bibitem{prnewswire2019}{
    ID2020 Alliance.
    ``ID2020 Alliance launches digital ID program with Government of Bangladesh and Gavi, announces new partners at annual summit.''
    PR Newswire,
    2019-09-19.
    [online]
    \url{https://www.prnewswire.com/news-releases/id2020-alliance-launches-digital-id-program-with-government-of-bangladesh-and-}
    \url{gavi-announces-new-partners-at-annual-summit-300921926.html}
    [retrieved 2019-10-11]
}
\bibitem{uport}{
    Lundkvist, C, Heck, R, Torstensson, J, Mitton, Z, and Sena, M.
    ``uPort: A Platform for Self-Sovereign Identity.''
    Draft Version, 2016-10-20.
    [online]
    \url{http://blockchainlab.com/pdf/uPort_whitepaper_DRAFT20161020.pdf}
    [retrieved 2019-01-11]
}
\bibitem{kaaniche2017}{
    Kaaniche, N and Laurent, M.
    ``A blockchain-based data usage auditing architecture with enhanced privacy and availability.''
    IEEE 16th International Symposium on Network Computing and Applications (NCA),
    2017-10-30.
    [online]
    \url{https://ieeexplore.ieee.org/abstract/document/8171384/}
    [retrieved 2019-01-11]
}
\bibitem{chainspace}{
    Chainspace.
    [online]
    \url{https://chainspace.io/}
    [retrieved 2019-10-11]
}
\bibitem{sonnino2018}{
    Sonnino, A et al.
    ``Coconut: Threshold Issuance Selective Disclosure Credentials with Applications to Distributed Ledgers.''
    arXiv:1802.07344v3 [cs.CR],
    2018-08-08.
    [online]
    \url{https://arxiv.org/abs/1802.07344}
    [retrieved 2019-01-11]
}
\bibitem{heath2019}{
    Heath, A.
    ``Facebook Makes First Blockchain Acquisition With Chainspace: Sources.''
    Cheddar, Inc,
    2019-02-04.
    [online]
    \url{https://cheddar.com/media/facebook-blockchain-acquisition-chainspace}
    [retrieved 2019-02-05]
}
\bibitem{field2019}{
    Field, M.
    ``The tiny UK start-up founded by UCL scientists now at the heart of Facebook's Libra currency.''
    \textit{The Telegraph},
    2019-06-27.
    [online]
    \url{https://www.telegraph.co.uk/technology/2019/06/26/inside-tiny-london-start-up-heart-facebooks-push-reinvent-world/}
    [retrieved 2019-10-11]
}
\bibitem{itu2018}{
    International Telecommunications Union.
    \textit{Digital Identity Roadmap Guide},
    2018.
    [online]
    \url{http://handle.itu.int/11.1002/pub/81215cb9-en}
    [retrieved 2019-10-01]
}
\bibitem{un1948}{
    United Nations.
    \textit{Universal Declaration of Human Rights}.
    General Assembly Resolution 217 A,
    Article 12.
    Paris,
    1948-12-10.
    [online]
    \url{http://www.ohchr.org/EN/UDHR/Documents/UDHR_Translations/eng.pdf}
    [retrieved 2019-10-11]
}
\bibitem{armer1975}{
    P. Armer.
    ``Computer Technology and Surveillance.''
    \textit{Computers and People} 24(9),
    pp. 8--11,
    September 1975.
    [online]
    \url{https://archive.org/stream/bitsavers_computersA_3986915/197509#page/n7/mode/2up}
    [retrieved 2017-05-28]
}
\bibitem{stoa1999}{
    European Parliament.
    ``Development of Surveillance Technology and Risk of Abuse of Economic Information.''
    \textit{Scientific and Technological Options Assessment (STOA)},
    PE 168.184/Vol 1/5/EN.
    Luxembourg,
    December 1999.
    [online]
    \url{www.europarl.europa.eu/RegData/etudes/etudes/join/1999/168184/DG-4-JOIN_ET%281999%29168184_EN.pdf}
    [retrieved 2019-10-11]
}
\bibitem{goodell18}{
    Goodell, G and Aste, T.
    ``Blockchain technology for the public good: Design constraints in a human rights context.''
    Open Access Government,
    May 2018.
    [online]
    \url{https://www.openaccessgovernment.org/blockchain-technology-for-the-public-good-design-constraints-in-a-human-rights-} \url{context/44595/}
}
\bibitem{mayo45}{
    Mayo, E.
    ``Hawthorne and the Western Electric Company.''
    In \textit{The Social Problems of an Industrial Civilization}.
    Boston: Division of Research, Harvard Business School,
    1945.
    [online]
    \url{https://web.archive.org/web/20181123151854/http://www.practicesurvival.com/wa_files/Hawthorne_20Studies_201924_20Elton_20Mayo.pdf}
    [retrieved 2019-01-06]
}
\bibitem{koslowska}{
    Koslowska, H, Gershgorn, D, and Todd, S.
    ``The Cambridge Analytica scandal is wildly confusing. This timeline will help.''
    \textit{Quartz},
    2018-03-29.
    [online]
    \url{https://qz.com/1240039/the-cambridge-analytica-scandal-is-confusing-this-timeline-will-help/}
    [retrieved 2018-04-20]
}
\bibitem{lunden}{
    Lunden, I.
    ``Russia’s Telegram ban that knocked out 15M Google, Amazon IP addresses had a precedent in Zello.''
    \textit{TechCrunch},
    2018-04-19.
    [online]
    \url{https://techcrunch.com/2018/04/17/russias-telegram-ban-that-knocked-out-15m-google-amazon-ip-addresses-had-}
    \url{a-precedent-in-zello/}
    [retrieved 2018-04-20]
}
\bibitem{savov}{
    Savov, V.
    ``Russia’s Telegram ban is a big, convoluted mess.''
    \textit{The Verge},
    2018-04-17.
    [online]
    \url{https://www.theverge.com/2018/4/17/17246150/telegram-russia-ban}
    [retrieved 2018-04-20]
}
\bibitem{riley}{
    Riley, S.
    ``It’s Me, and Here’s My Proof: Why Identity and Authentication Must Remain Distinct.''
    Microsoft TechNet Security Viewpoint,
    2006-02-14.
    [online]
    \url{https://technet.microsoft.com/en-us/library/cc512578.aspx}
    [retrieved 2018-05-11]
}
\bibitem{ibm1}{
    IBM Research Zurich.
    ``IBM Identity Mixer.''
    [online]
    \url{https://www.zurich.ibm.com/identity_mixer/}
    [retrieved 2018-04-21]
}
\bibitem{camenisch2003}{
    Camenisch, J and Lysyanskaya, A.
    ``A Signature Scheme with Efficient Protocols.''
    \textit{Lecture Notes in Computer Science},
    vol. 2576.
    Springer,
    2003.
    [online]
    \url{http://rd.springer.com/chapter/10.1007%2F3-540-36413-7_20}
    [retreived 2017-05-10]
}
\bibitem{skc}{
    SecureKey Technologies, Inc.
    ``Trust Framework -- SecureKey Concierge in Canada, SKUN-117.''
    2015-09-09.
    [online]
    \url{http://securekey.com/wp-content/uploads/2015/09/SK-UN117-Trust-Framework-SecureKey-Concierge-Canada.pdf}
    [retrieved 2018-05-11]
}
\bibitem{trossen2005}{
    Value Chain Dynamics Working Group (VCDWG).
    ``Value Chain Dynamics in the Communication Industry.''
    MIT Communications Futures Program and Cambridge University Communications Research Network,
    December 2005.
    [online]
    \url{http://cfp.mit.edu/docs/core-edge-dec2005.pdf}
    [retrieved 2019-01-11]
}
\bibitem{ramakrishnan2017}{
    Ramakrishnan, K and Selvarajan, R.
    ``Transforming the Telecom Value Chain with Platformization.''
    TATA Consultancy Services White Paper,
    2017.
    [online]
    \url{https://www.tcs.com/content/dam/tcs/pdf/Industries/communication-media-and-technology/Abstract/Transforming%20the%20Telecom%20Value%20Chain%20with%20Platformization.pdf}
    [retrieved 2019-01-11]
}
\bibitem{debian}{
    Software in the Public Interest, Inc.
    ``Debian: The Universal Operating System.''
    [online]
    \url{https://www.debian.org/}
    [retrieved 2019-10-11]
}
\bibitem{arduino}{
    Arduino.
    [online]
    \url{https://www.arduino.cc/}
    [retrieved 2019-10-11]
}
\bibitem{govuk2014}{
    GOV.UK.
    ``Money Laundering Regulations: who needs to register.''
    2014-10-23.
    [online]
    \url{https://www.gov.uk/guidance/money-laundering-regulations-who-needs-to-register}
    [retrieved 2017-05-28]
}
\bibitem{amlkyc}{
    Better Business Finance.
    ``What are the AML and KYC obligations of a Bank in the UK?''
    [online]
    \url{https://www.betterbusinessfinance.co.uk/aml-and-kyc/what-are-the-aml-and-kyc-obligations-of-a-bank-in-the-uk}
    [retrieved 2017-05-28]
}
\bibitem{chaum82}{
    Chaum, D.
    ``Blind signatures for untraceable payments.''
    In \textit{Advances of Cryptology Proceedings of Crypto}.
    \textbf{82}(3), pp. 199--203,
    1983.
    \url{http://www.hit.bme.hu/~buttyan/courses/BMEVIHIM219/2009/Chaum.BlindSigForPayment.1982.PDF}
    [retrieved 2018-02-23]
}
\bibitem{ocsp}{
    Juniper Networks.
    ``Understanding Online Certificate Status Protocol.''
    2018-02-27.
    [online]
    \url{https://web.archive.org/web/20191024205445/https://www.juniper.net/documentation/en_US/junos/topics/concept/certificate-ocsp-understanding.html}
    [retrieved 2018-05-11]
}
\bibitem{swift}{
    Society for Worldwide Interbank Financial Telecommunication.
    ``Discover SWIFT.''
    [online]
    \url{https://www.swift.com/about-us/discover-swift}
    [retrieved 2018-04-21]
}
\bibitem{x12}{
    The Accredited Standards Committee.
    ``ASC X12.''
    [online]
    \url{https://web.archive.org/web/20140927153741/http://www.x12.org/x12org/about/index.cfm}
    [retrieved 2018-04-21]
}
\bibitem{edifact}{
    United Nations Economic Commission for Europe.
    ``Trade Programme -- Trade -- UNECE.''
    [online]
    \url{http://www.unece.org/tradewelcome/trade-programme.html}
    [retrieved 2018-04-21]
}
\bibitem{ripple}{
    Ripple: One Frictionless Experience To Send Money Globally.
    [online]
    \url{https://ripple.com/}
    [retrieved 2019-01-11]
}
\bibitem{hyperledger}{
    IBM.
    ``Hyperledger: blockchain collaboration changing the business world.''
    [online]
    \url{https://web.archive.org/web/20190803212916/https://www.ibm.com/blockchain/hyperledger}
    [retrieved 2019-01-11]
}
\bibitem{regnms}{
    Securities and Exchange Commission (US).
    \textit{REGULATION NMS.}
    17 CFR PARTS 200, 201, 230, 240, 242, 249, and 270;
    Release No. 34-51808;
    File No. S7-10-04,
    RIN 3235-AJ18,
    2005-06-09.
    [online]
    \url{https://www.sec.gov/rules/final/34-51808.pdf}
    [retrieved 2019-01-30]
}
\bibitem{mifid}{
    European Parliament.
    Directive 2004/39/EC.
    \textit{Official Journal of the European Union},
    2004-04-21.
    [online]
    \url{https://eur-lex.europa.eu/legal-content/EN/ALL/?uri=CELEX:02004L0039-20060428}
    [retrieved 2019-10-01]
}
\bibitem{regnms-timeline}{
    Securities and Exchange Commission Historical Society.
    ``2000s: Timeline.''
    [online]
    \url{https://web.archive.org/web/20190309003627/http://www.sechistorical.org/museum/timeline/2000-timeline.php}
    [retrieved 2018-04-19]
}
\bibitem{ice}{
    Reuters.
    ``ICE completes takeover of NYSE.''
    [online]
    \url{https://www.reuters.com/article/us-ice-nyse-sprecher/ice-completes-takeover-of-nyse-idUSBRE9AB16V20131112}
    [retreived 2018-04-19]
}
\bibitem{zimmer2018}{
    Zimmer, M and Tijerina, B.
    ``Library Values \& Privacy in our National Digital Strategies: Field guides, Convenings, and Conversations.''
    National Leadership Grant for Libraries Award Report.
    Milwaukee, Wisconsin, USA,
    2018.
    [online]
    \url{https://www.michaelzimmer.org/2018/08/02/project-report-library-values-privacy/}
    [retrieved 2019-10-05]
}
\bibitem{heydt2006}{
    Heydt-Benjamin, T, Chae, H, Defend, B, and Fu, K.
    ``Privacy for Public Transportation.''
    International Workshop on Privacy Enhancing Technologies (PETS),
    2006,
    pp. 1--19.
    [online]
    \url{https://petsymposium.org/2006/preproc/preproc_01.pdf}
    [retrieved 2019-10-05]
}
\bibitem{imei}{
    GSM Association.
    ``IMEI Database.''
    [online]
    \url{https://imeidb.gsma.com/imei/index}
    [retrieved 2019-10-11]
}
\bibitem{goodell2019}{
    Goodell, G and Aste, T.
    ``Can Cryptocurrencies Preserve Privacy and Comply with Regulations?''
    \textit{Frontiers in Blockchain},
    May 2019.
    \url{doi:10.3389/fbloc.2019.00004}.
    Also available at SSRN: \url{https://ssrn.com/abstract=3293910}.
}

\end{thebibliography}
\end{document}